\documentclass[10pt,journal,twocolumn]{IEEEtran} 
\usepackage{wrapfig,booktabs,fancyhdr,amsmath,amsfonts,tabularx,numprint}
\usepackage{cite,bm,bbm,amssymb,amsthm,url,multirow,times,enumitem,comment}
\usepackage{mathtools,siunitx,balance,tikz,adjustbox,graphicx,array,listings}
\usepackage{xspace,mathrsfs}
\usepackage{adjustbox}
\usepackage[font=footnotesize]{subcaption}
\usepackage[linesnumbered,ruled]{algorithm2e}
\usepackage[colorlinks=true, linkcolor=black, citecolor=blue, urlcolor=blue]{hyperref}
\usepackage[utf8]{inputenc}
\usepackage[T1]{fontenc}

\newcommand{\vb}{\boldsymbol}
\newcommand{\vbh}[1]{\hat{\boldsymbol{#1}}}

\newcommand{\abs}[1]{\left|{#1}\right|}

\newcommand{\ba}{\begin{array}}
\newcommand{\ea}{\end{array}}

\newcommand{\define}{\triangleq}

\newcommand{\E}[1]{\mathbb{E}\left[ #1 \right]}

\DeclareMathAlphabet{\mathpzc}{OT1}{pzc}{m}{it}
\DeclareMathOperator*{\argmin}{argmin}
\DeclareMathOperator*{\argmax}{argmax}
\usetikzlibrary{shapes,arrows.meta,positioning,calc,patterns.meta}
\tikzstyle{block} = [draw,fill=white,rectangle,thick,minimum height=3em,minimum width=6em]
\tikzstyle{sum} = [draw, fill=white, circle, node distance=1cm]
\tikzstyle{connector} = [-Stealth, thick]
\tikzstyle{dotconnector} = [{Circle[length=4pt]}-Stealth, thick, shorten <=-2pt]
\tikzstyle{input} = [coordinate]
\tikzstyle{output} = [coordinate]

\newcommand{\C}{\mathcal{C}}
\newcommand{\CC}[1]{\mathcal{C}_{#1}}
\newcommand{\hCC}[1]{\hat{\mathcal{C}}_{#1}}

\newcommand{\Cqpsk}{\mathcal{C}^\text{QPSK}}
\newcommand{\Cfqam}{\mathcal{C}^\text{4QAM}}
\newcommand{\Cstqam}{\mathcal{C}^\text{16QAM}}
\newcommand{\Cttqam}{\mathcal{C}^\text{32QAM}}

\newcommand{\xlr}{x_{m01}}
\newcommand{\ylr}{y_{01}}
\newcommand{\tylr}{\tilde{y}_{01}}

\newcommand{\Ts}{T_\text{s}}

\newcommand{\Fc}[1]{F_{\text{c}#1}}
\newcommand{\Fs}{F_\text{s}}

\newcommand{\Tsym}{T_\text{sym}}

\newcommand{\Tf}{T_\text{f}}

\newcommand{\Nsf}{N_\text{sf}}
\newcommand{\Ns}{N_\text{s}}

\newcommand{\F}{F}
\newcommand{\Tg}{T_\text{g}}
\newcommand{\Ng}{N_\text{g}}

\newcommand{\vlos}{v_\text{los}}

\newcommand{\Tfg}{T_\text{fg}}

\newcommand{\Mod}{\:\mathrm{mod}\:}

\newcommand{\td}{t_\text{d}}

\newcommand{\tf}{t_\text{f}}
\newcommand{\tc}{t_\text{c}}
\newcommand{\ts}{t_\text{s}}

\newcommand{\dtf}{\delta\tf}
\newcommand{\dtc}{{\delta\tc}}
\newcommand{\dts}{{\delta\ts}}

\newcommand{\dtcd}{\dot{\dtc}}
\newcommand{\dtsd}{\dot{\dts}}

\newcommand{\dtcn}{\delta t_{\text{c}0}}
\newcommand{\dtsn}{\delta t_{\text{s}0}}

\newcommand{\betas}{\beta_{\text{s}}}
\newcommand{\dbetas}{\delta\betas}

\newcommand{\betac}{\beta_{\text{c}}}
\newcommand{\dbetac}{\delta\betac}

\newcommand{\X}{\boldsymbol{X}}
\newcommand{\D}{\boldsymbol{D}}

\newcommand{\hXm}{\hat{\boldsymbol{X}}_m}

\newcommand{\hXmi}{\hat{\boldsymbol{X}}_{mi}}
\newcommand{\vbY}[1]{\boldsymbol{Y}_{#1}}
\newcommand{\bYmi}{\bar{\boldsymbol{Y}}_{mi}}
\newcommand{\tYmi}{\tilde{\boldsymbol{Y}}_{mi}}

\newcommand{\SNRpost}{\text{SNR}_\text{post}}
\newcommand{\SNRpre}{\text{SNR}_\text{pre}}

\newcommand{\K}{\mathcal{K}}
\newcommand{\Kl}{\mathcal{K}_\text{l}}
\newcommand{\Kp}{\mathcal{K}_\text{p}}
\newcommand{\Klnp}{\mathcal{K}_\text{lnp}}
\newcommand{\Kg}{\mathcal{K}_\text{g}}

\newcommand{\Ids}{\mathcal{I}}
\newcommand{\Idm}{\tilde{\mathcal{I}}_{m}}
\newcommand{\Ido}{\mathcal{I}_1}
\newcommand{\Idt}{\mathcal{I}_2}
\newcommand{\IQm}{\mathcal{I}_{\text{Q}m}}
\newcommand{\IQmn}{\mathcal{I}_{\text{Q}mn}}
\newcommand{\ITm}{\mathcal{I}_{\text{T}m}}
\newcommand{\IHm}{\mathcal{I}_{\text{H}m}}
\newcommand{\Tref}{\vb{T}}
\newcommand{\dT}{d_\text{T}}
\newcommand{\ihm}{i_{\text{h}m}}
\newcommand{\spik}{s_{\text{p}ik}}
\newcommand{\qpk}{q_{\text{p}k}}
\newcommand{\Mex}{\mathcal{M}_\text{e}}
\newcommand{\Mexp}{\mathcal{M}_\text{ep}}
\newcommand{\NT}{N_\text{T}}

\begin{document}
\typeout{COLUMNWIDTH: \the\columnwidth}
\typeout{TEXTWIDTH: \the\textwidth}
\title{Pilots and Other Predictable Elements of the Starlink Ku-Band Downlink}

\author{
  \IEEEauthorblockN{Wenkai Qin\IEEEauthorrefmark{1},
                    Mark L. Psiaki\IEEEauthorrefmark{2},
                    John R. Bowman\IEEEauthorrefmark{2},
                    Todd E. Humphreys\IEEEauthorrefmark{1}} \\
  \IEEEauthorblockA{\IEEEauthorrefmark{1}\textit{Department of Aerospace Engineering and Engineering Mechanics, The University of Texas at Austin}} \\
  \IEEEauthorblockA{\IEEEauthorrefmark{2}\textit{Department of Aerospace and Ocean Engineering, Virginia Tech}}
}

\maketitle

\begin{abstract}
  We identify and characterize dedicated pilot symbols and other predictable
  elements embedded within the Starlink Ku-band downlink waveform.  Exploitation
  of these predictable elements enables precise opportunistic positioning,
  navigation, and timing using compact, low-gain receivers by maximizing the
  signal processing gain available for signal acquisition and time-of-arrival
  (TOA) estimation. We develop an acquisition and demodulation framework to
  decode Starlink frames and disclose the explicit sequences of the \textit{edge
    pilots}---bands of 4QAM symbols located at both edges of each Starlink
  channel that apparently repeat identically across all frames, beams, channels,
  and satellites. We further reveal that the great majority of QPSK-modulated
  symbols do not carry high-entropy user data but instead follow a regular
  tessellated structure superimposed on a constant reference template. We
  demonstrate that exploiting frame-level predictable elements yields a
  processing gain of approximately 48 dB, thereby enabling low-cost, compact
  receivers to extract precise TOA measurements even from low-SNR Starlink side
  beams.
\end{abstract}


\newif\ifpreprint
\preprintfalse

\ifpreprint

\pagestyle{plain}
\thispagestyle{fancy}  
\fancyhf{} 
\renewcommand{\headrulewidth}{0pt}
\rfoot{\footnotesize \bf October 2025 preprint of paper submitted for review} \lfoot{\footnotesize \bf
  Copyright \copyright~2025 by Wenkai Qin \\ and Todd E. Humphreys}

\else

\thispagestyle{empty}
\pagestyle{plain}   

\fi


\section{Introduction}
\label{sec:introduction}
Position, navigation, and timing (PNT) is a crucial utility whose applications
span vast sectors of modern life.  From agriculture to aviation to financial
markets, Global Navigation Satellite Systems (GNSS)---the prevailing technology
for PNT---provide the nanosecond-level timing and meter-level positioning
necessary for efficient logistics, safe transportation, synchronized
telecommunications, and precise financial transactions.  Yet legacy GNSS face
both a set of critical vulnerabilities and an expanding number of threats.
Because legacy civil GNSS transmit weak L-band signals from medium Earth orbit
and operate as open-access systems, users risk denial of service and deception
by jamming and spoofing attacks.  These risks are intensifying as the space
domain and electromagnetic spectrum emerge as decisive arenas of 21st-century
warfare, with predictable and escalating dangers for civilian users of
GNSS-based PNT services \cite{mcneil2025gpsjamming, zulhusni2025gpsinterference}.

Meanwhile, advances in reusable rocket technology have led to a tenfold
reduction in space launch costs, laying the groundwork for a rapidly expanding
low Earth orbit (LEO) space economy \cite{jones2018recent}.  Recently launched
LEO communications systems (e.g., OneWeb, Starlink, and Amazon Leo) have already
attracted millions of subscribers and currently generate billions of dollars in
annual revenue.  These systems have drawn the interest of PNT researchers
seeking a robust, precise, and broadly accessible alternative to legacy GNSS:
the consistent availability, high power, and wide bandwidth of LEO
communications signals confer inherent resilience to adversarial interference
and great potential for PNT precision.  But despite the significant potential
and low marginal cost of fusing a PNT service with an existing broadband LEO
communications service \cite{iannucci2020fused}, no major constellation
operators have announced plans to bring forward a PNT service whose accuracy
would match or exceed that of legacy GNSS.  Meanwhile, researchers have
developed techniques to opportunistically exploit LEO communications signals for
PNT, with efforts from several groups already bearing fruit
\cite{tan2019positioning, lei2025robust, kassas2024adAstra,
  kozhaya2025unveiling, jardak2023practical, neinavaie2023cognitive,
  grayver2024position, morgan2025mockplans, komodromos2025networkplans}.  While
many constellations' signals have been shown to be exploitable
\cite{tan2019positioning, komodromos2025networkplans, kassas2024adAstra},
Starlink remains the most attractive candidate for opportunistic navigation:
with a constellation of over 10,000 satellites, Starlink's dense, geometry-rich
coverage reaches nearly the entire globe and offers a promising foundation for
multilaterated PNT \cite{iannucci2022fusedleo}.

\begin{figure}[t]
  \centering
  \begin{subfigure}[t]{0.68\columnwidth}
    \centering
    \includegraphics[trim=10 0 20 10,clip,width=\linewidth]{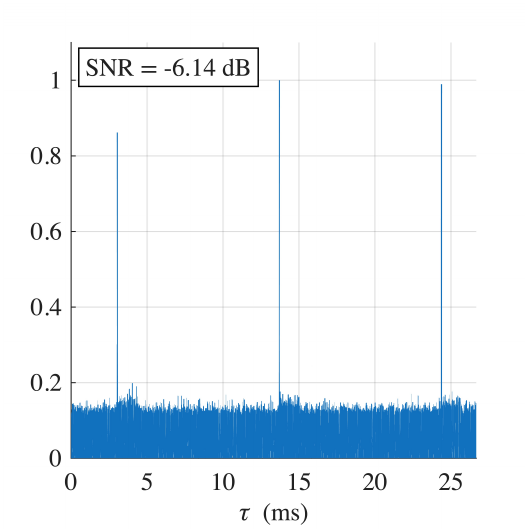}
  \end{subfigure}
  \begin{subfigure}[t]{0.23\columnwidth}
    \centering
    \raisebox{0.08\textheight}{
    \includegraphics[width=\linewidth]{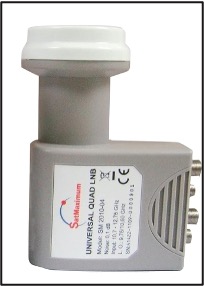}
    }
  \end{subfigure}

  \hfill
  \caption{Normalized matched-filter correlation against a coherent combination
    of the known primary and secondary synchronization sequences (PSS and SSS)
    with a signal captured from a compact feedhorn receiver of the type
    shown. This receiver, intended for consumer-grade satellite television
    reception in the \qtyrange{10.7}{12.75}{\giga\hertz} frequency band,
    provides a \qtyrange{50}{60}{\decibel} gain and a noise figure below
    \qty{1}{\decibel}. The SNR value shown indicates the pre-correlation SNR of
    the received signal associated with the strongest post-correlation peak. For
    this capture, which spans the full \qty{240}{\mega\hertz} bandwidth of a
    single Starlink channel, the theoretical processing gain by which the
    post-correlation SNR exceeds the pre-correlation SNR is \qty{33}{\decibel}.}
  \label{fig:cfr-acq-snr}
\end{figure}

Nonetheless, several challenges remain before Starlink-based PNT can be a
practically viable alternative to legacy GNSS. One major hurdle concerns the
physical footprint of effective Starlink PNT receiver antennas.  Supposing
Starlink phased array terminal form-factors are indicative of a future
PNT-capable terminal, even the smallest option currently offered by SpaceX
(Starlink Mini) measures \qtyproduct{30 x 26}{\cm}, which is too large for many
practical applications (e.g., small unmanned aerial vehicles, consumer
electronics, and asset trackers).  A compact, mass-produced low-noise block
(LNB) with an integrated hemispherical feedhorn, such as the unit shown in
Fig.~\ref{fig:cfr-acq-snr}, offers a more attractive physical profile, with
feedhorn outer diameter \qty{6}{\cm}, maximal dimension \qty{12}{\cm}, and mass
under \qty{200}{\gram}.

The use of such a compact feedhorn receiver for Starlink-based PNT would confer
an additional benefit: its wide beamwidth enables simultaneous reception of
multiple satellite transmissions, which allows multilateration, provided that
each transmitting satellite can be uniquely associated with its corresponding
component in the received signal.  Indeed, the systems presented in
\cite{neinavaie2021acquisition,jardak2023practical,kozhaya2023starlinkdoppler}
exclusively employ compact feedhorn receivers to track so-called pilot tones for
Starlink-based Doppler positioning. Unfortunately, these pilot tones' strength
and availability have significantly decreased since 2023, rendering this
approach ineffective \cite{kozhaya2025unveiling}.

In \cite{grayver2024position}, researchers describe a pseudorange-based Starlink
PNT system with an approximately \qty[quantity-product={-}]{20}{\centi\meter}
feedhorn (yielding ~\qty{15}{\decibel i} of gain) coupled to a wideband LNB.
However, the resulting position solutions remain relatively coarse, with miss
distances on the order of \qty{100}{\meter} or more.  The system presented in
\cite{morgan2025mockplans} geolocates a compact feedhorn receiver, but it relies
on a stationary reference receiver that relays high-bitrate data captured using
a \qty[quantity-product={-}]{35}{\decibel i},
\qty[quantity-product={-}]{90}{\centi\meter} parabolic dish antenna to aid
signal tracking and disambiguation.

Thus, although the feasibility of Starlink-based PNT has been repeatedly
demonstrated, achieving precise PNT with a compact, low-cost antenna and
minimal side-channel aiding remains a challenge.

Ultimately, the practicality and performance of a Starlink-based PNT system are
subject to constraints on the receiver's antenna gain, which must provide a
sufficient signal-to-noise ratio (SNR) to ensure reliable signal acquisition and
satisfactory PNT precision. For instance, consider a Ku-band horn antenna with a
circular \qty{54}{\milli\meter} diameter aperture. Supposing an operating
frequency of \qty{12}{\giga\hertz} and antenna efficiency of $\eta = 0.7$, this
antenna's gain is theoretically limited to approximately \qty{15.1}{\decibel i}
with an \qty{32}{\degree} half-power beamwidth, as calculated using standard
antenna theory \cite[Section~2-11]{johnson1993antenna},
\cite[Eq.~(12-29a)]{balanis2005antennatheory}. Assuming clear-sky conditions, an
LNB noise figure of \qty{1}{\decibel}, and published Starlink Ku-band power flux
density (PFD) of $-145$ to \qty{-138}{\decibel W} per \qty{4}{\kilo\hertz}, this
antenna yields a maximum pre-correlation SNR in the range of \qty{1}{\decibel}
to \qty{8}{\decibel} \cite[Sec.~2.16--2.18]{balanis2005antennatheory},
\cite{nrao2023evlamemo}.

Our own measurements of Starlink pre-correlation SNR from an assigned beam
signal (one directed to the receiver's service cell) captured through a compact
feedhorn antenna with an integrated LNB lie several decibels below this range,
as shown in Fig.~\ref{fig:cfr-acq-snr}.  This is because, in practice, SNR is
degraded by off-axis reception of the satellite beam, polarization mismatch,
receiver implementation losses, atmospheric attenuation, and co-channel
interference from other beams and satellites.

Nonetheless, the clearly defined correlation peaks in Fig.~\ref{fig:cfr-acq-snr}
show that matched filtering of an assigned beam's signal against a local replica
composed of the coherent combination of the primary and secondary
synchronization sequences (PSS and SSS---described in
\cite{humphreys2023starlinksignalstructure}) can provide ample processing gain
for signal acquisition and tracking, assuming the signal is captured with
adequate bandwidth and under low-interference conditions. But this result does
not necessarily extend to side beams---those directed to a service cell other
than the one in which the receiver resides.  The authors of
\cite{qin2025starlinktimingproperties} observe that a composite signal's
trackable side beams may have SNRs that are $5$ to \qty{18}{\decibel} below that
of the assigned beam---too low to extract useful time-of-arrival (TOA)
measurements solely with the PSS+SSS local replica, as will be shown in
Section~\ref{sec:toa-snr}.  The ability to extract sub-meter-level-precise TOA
measurements from several side beams would improve geometric dilution of
precision (GDOP) and shorten time-to-first-fix.

The key to unlocking robust and practical Starlink-based PNT with a compact
receiver is to boost the processing gain well beyond that provided by the
PSS+SSS combination.  Processing gain, a fundamental concept in direct-sequence
spread-spectrum (DSSS) systems such as code-division multiple access (CDMA) and
GNSS, can be extended to opportunistic exploitation of orthogonal frequency
division multiplexing (OFDM) waveforms such as Starlink's, in which context the
processing gain scales as the number of known information symbols.  With
adequate processing gain, the side beam signals required for a fast and precise
Starlink PNT solution can achieve sufficient post-correlation SNR to permit
precise TOA measurement.

We note that, in principle, an opportunistic PNT user could apply the
maximum-likelihood (ML) or decision-directed (DD) methods of
\cite{graff2025data} to recover TOA measurements despite lack of knowledge of
the OFDM information symbols. But these methods break down catastrophically in
the low-SNR regime even for low-order modulation (see Fig. 1 in
\cite{graff2025data}), and so are inapt for Starlink signals received through
compact antennas.

While known information symbols are the most straightforward features to exploit
\textit{a priori}, knowledge of any invariant property of the Starlink frame
likewise increases the processing gain---for instance, the range of modulation
types, the temporal arrangement of modulation, or whether individual OFDM
symbols are simple in their modulation (e.g., purely QPSK) or composite (e.g., a
mixture of QPSK and 16QAM).  Accordingly, we aim to identify key predictable
features present in the Starlink Ku-band downlink waveform, thereby boosting the
achievable processing gain.

Prior work already discovered several predictable features.  Reference
\cite{humphreys2023starlinksignalstructure} revealed the Starlink frame
structure, including the PSS and SSS and their explicit values.  SpaceX's 2024
patent \cite{mccormick2024ofdm} confirmed this frame structure and additionally
disclosed the existence of two bands of pilot symbols near the edges of each
Starlink channel's allocated bandwidth.  Pilot symbols are complex-valued
coefficients that modulate OFDM subcarriers just like other information symbols,
but are entirely predictable to the receiver and so can be used for channel
estimation.  Reference \cite{kozhaya2025unveiling} empirically confirmed these
pilot bands, hereafter called the edge pilots, and reported additional
inter-frame-correlated data not disclosed by \cite{mccormick2024ofdm}, but
implied that these additional correlated data appear in all frames.  This, as we
explain herein, does not appear to be true.  Moreover,
\cite{kozhaya2025unveiling} did not provide the actual information symbol values
for the edge pilots or any other inter-frame-correlated data, and so does not
enable the most potent method for signal detection, namely, matched filtering
against a noise-free local signal replica.

The purpose of the current paper is to offer the fullest account yet of
predictable structure in the Starlink waveform.  Not only do we reveal the exact
information symbols of the edge pilots, but we also report a remarkable
discovery: after the first few OFDM symbols in each frame, which appear to
contain header information, almost all symbols modulated as QPSK are nearly
perfectly predictable and thus low-entropy.  While this fact is not obvious from
a cursory examination of the data, application of certain transformations
described in detail herein reveals a highly consistent structure with extended
segments of predictable symbols.  Fully exploiting these low-entropy signal
elements allows matched filtering with sufficient processing gain to support
robust PNT with low-cost compact feedhorn receivers.

\section{SNR-Dependent TOA Precision}
\label{sec:toa-snr}

\begin{figure}[t]
  \centering
  \includegraphics[width=\columnwidth]{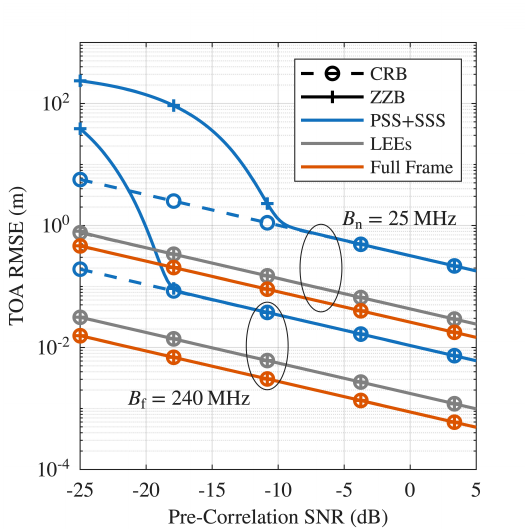}
  \caption{Lower bounds on single-frame TOA root mean squared error (RMSE)
    vs. pre-correlation SNR for coherent processing with three representative
    local replicas: (1) the PSS+SSS combination (blue), (2) low-entropy elements
    (LEEs) of the frame (gray), and (3) the full Starlink frame (red).  Results
    are shown for two representative capture bandwidths, $25$ and
    $\qty{240}{\mega\hertz}$.  The Ziv-Zakai bounds (ZZBs; solid lines) are
    tighter than the Cramér-Rao bounds (CRBs; dashed lines) at low SNR because
    they capture the threshold effects that lead to rapid degradation in TOA
    precision in that regime.  The SNR range shown covers the relevant values
    for trackable signals from assigned beams (high SNR) to side beams (low
    SNR).  The bounds based on LEEs are discussed further in
    Section~\ref{sec:known-predictable}.}
  \label{fig:zz-cr-bounds}
\end{figure}

To assess the conditions under which precise TOA estimation is feasible, we
examine the Ziv-Zakai Bound (ZZB), which characterizes estimator performance
across a given SNR range by explicitly modeling the transition from small-error
behavior in the high-SNR regime, where the Cramér-Rao Bound (CRB) is
asymptotically tight, to large-error behavior dominated by noise and
ambiguity. In contrast to the CRB, which assumes unbiased estimation in the
high-SNR limit, the ZZB captures threshold effects that lead to rapid
degradation in TOA precision at low SNR.  Reference \cite{graff2024ziv}
describes in detail the ZZB formulation and its application to OFDM-based TOA
estimation.

Fig.~\ref{fig:zz-cr-bounds} shows TOA precision bounds as functions of
pre-correlation SNR for representative Starlink processing strategies.  When
correlating against the relatively short PSS+SSS local replica using the
full-channel bandwidth $B_\text{f}=\qty{240}{\mega\hertz}$, the ZZB departs from
the CRB for pre-correlation SNR below approximately
\qty{-17.4}{\decibel}. Narrowing the capture bandwidth to
$B_\text{n}=\qty{25}{\mega\hertz}$ shifts this divergence point to a
substantially higher SNR of approximately \qty{-9.2}{\decibel}. Below these
thresholds, TOA precision degrades rapidly.

In contrast, when correlating against a full-frame local replica---one that
spans all subcarriers and OFDM symbols of the Starlink frame---the substantially
increased number of coherently accumulated samples yields a greater processing
gain and suppresses ambiguity sidelobes in the correlation function. As a
result, the ZZB remains coincident with the CRB across the entire SNR range
shown, even when the capture bandwidth is limited to $B_\text{n}$.  These results
indicate that full-frame processing maintains CRB-like TOA estimation over a
wider SNR range than PSS+SSS-only processing.

One can think of the PSS+SSS-only and full-frame strategies as extreme cases
that bookend reasonable coherent processing for TOA measurements from Starlink
signals: There is little reason to process less than the coherent combination of
the PSS and SSS, which are known and easily constructed
\cite{humphreys2023starlinksignalstructure}, whereas coherent processing beyond
one full frame is complicated by inter-frame carrier phase discontinuities,
which, so far, have resisted modeling.  Thus, Fig.~\ref{fig:zz-cr-bounds} covers
all reasonable cases between short-duration processing with narrow-bandwidth
captures (the PSS+SSS case with $B_\text{n}$) and full-frame processing with
full-channel-bandwidth captures (the full-frame case with $B_\text{f}$).

Fig.~\ref{fig:zz-cr-bounds} makes clear that a receiver seeking to suppress the
ZZB-predicted breakdown in TOA precision at low SNR must adopt strategies that
provide greater processing gain than that afforded by a PSS+SSS matched filter.
Such strategies extend the SNR range over which TOA estimation remains
asymptotic and enable the extraction of accurate TOA measurements from weaker
side beams that would otherwise be unusable.

\section{Processing Gain}
\label{sec:processing-gain}
Here we define processing gain, extend it to OFDM, and show how any invariant
property of the Starlink frame may be exploited to increase it.

For any signal and processing strategy, define the processing gain as
\begin{equation}
  \label{eq:Ldef}
  L \define \frac{\SNRpost}{\SNRpre}
\end{equation}
where $\SNRpre$ and $\SNRpost$ denote pre- and post-correlation SNR,
respectively. The former is the ratio of the total signal power to the total
noise power within the signal bandwidth. The latter is the ratio of the
signal-to-noise power within the de-spread bandwidth, or, equivalently, the SNR
of the complex product that results from correlation against whatever local
replica is used for detection. For DSSS despreading, this definition recovers
the familiar relation $L = B_\text{spread}/B_\text{despread}$, where
$B_\text{spread}$ and $B_\text{despread}$ are the spread and despread
bandwidths, respectively.

Consider a sequence of complex information symbols subject to additive noise
\begin{equation}
  \label{eq:noisySamples}
  r_k = x_k + n_k, \quad k \in \{1, \dots, N\}
\end{equation}
where the $k$th information signal $x_k \in \C$ is drawn uniformly from the
modulation constellation $\C = \{c^{(1)}, c^{(2)}, \dots, c^{(M)}\}$ and the
symbols $\{x_k \}_{k=1}^N$ are statistically independent. The constellation is
assumed to be normalized and zero-mean such that $\E{\lvert x_k \rvert^2} = 1$
and $\E{x_k}=0$ for all $k \in \{1, \dots, N\}$. The additive noise samples
$\{n_k\}_{k=1}^N$ are complex Gaussian random variables with
$n_k \sim \mathcal{CN}(0, \sigma_n^2)$, mutually independent for all
$k \in \{1, \dots, N\}$, and independent of $\{x_k \}_{k=1}^N$.

Define the signal detection statistic as the complex correlation product
\begin{align}
  \label{eq:detectionStat}
  S &\define \sum_{k=1}^{N} r_k^* l_k = \sum_{k=1}^{N} x_k^* l_k +
      \sum_{k=1}^{N} n_k^* l_k  = S_x + S_n
\end{align}
where $l_k$ is the $k$th element of a local replica sequence and $(\cdot)^*$
denotes the complex conjugate.  Pre- and post-correlation SNR can then be
defined as
\begin{align*}
  \SNRpre  \define \frac{\E{x_k^* x_k}}{\E{n_k^* n_k}} = \frac{1}{\sigma_n^2}, \quad
  \SNRpost \define \frac{\E{S_{x}^* S_{x}}}{\E{S_{n}^* S_{n}}}
\end{align*}

We additionally define the following correlation parameters for which
expectations are taken over the joint distribution of $x_k$ and $l_k$ across all
$k \in \{1, \dots, N\}$:
\begin{align}
  \label{eq:processGainCoeffs}
  \alpha \define \E{|x_k|^2 |l_k|^2}, \quad \rho \define \E{|l_k|^2}, \quad
  \mu \define \E{l_k^*\, x_k}
\end{align}
Now consider $L$ under each of three conditions for $\{x_k\}_{k=1}^N$: (i)
completely unknown, (ii) completely known, or (iii) partially known.  These
conditions can be modeled through the \emph{a priori} probability mass function
(PMF) of $x_k$ over $\C$.  In the unknown case, the PMF is uniform over $\C$.
In the known case, it collapses to a point mass at the transmitted symbol.  In
the partially known case, side information or invariant properties of the given
waveform may be exploited to restrict the PMF to a subset of the available
constellation points or otherwise bias it away from a uniform distribution.  In
this paper's
\href{https://rnl-data.ae.utexas.edu/datastore/supplementaryMaterial/qin-starlink-pilots/}{Supplementary
  Material} we evaluate $L$ under each condition and provide a general result.

For the special case of constant-modulus constellations (e.g., QPSK) and
unit-modulus replicas, we have $|x_k| = |l_k| = 1$, from which it follows that
$\alpha = \rho = 1$ and
\begin{equation}
  \label{eq:processing-gain}
  L = 1 + (N-1)|\mu|^2
\end{equation}
This result interpolates smoothly between the incoherent limit $L = 1$ for
completely unknown data ($|\mu| = 0$), and the fully coherent limit $L = N$ for
completely known data ($l_k = x_k$, $|\mu| = 1$).

Note that it makes no difference whether the complex information symbols in
\eqref{eq:noisySamples} represent samples in the time domain, as in DSSS, or the
frequency domain, as in OFDM.  Since the discrete Fourier transform (DFT) is a
linear, unitary transformation that preserves inner products and signal energy
(up to a constant normalization), the detection statistic
\eqref{eq:detectionStat} and its properties are preserved through the
transformation.  Consequently, correlating in the frequency domain across
subcarriers yields the same processing gain as correlating in the time domain
across samples.

In general, any invariant property of an OFDM waveform can be exploited during
the construction of $\{l_k\}_{k=1}^N$ to increase $|\mu|$ and thus $L$.  For
example, knowledge that certain OFDM symbols are simple rather than composite in
their modulation, or that their information symbols are merely a low-entropy
transformation of a known sequence, or knowledge of the guard (cyclic prefix;
CP) length $\Ng$, can be embedded in the index-dependent PMF for $x_k$ and thus
exploited to raise $L$.

For a generic OFDM waveform, calculation of $L$ is complicated by the CP, which
introduces correlation between the information symbols $\{x_k\}_{k=1}^N$,
whether viewed in the frequency or time domain.  For the particular case of the
Starlink Ku-band downlink waveform, calculation is further complicated by
time-domain correlation arising due to the repetitive structure of the PSS and
to the mid-channel gutter in which four adjacent subcarriers have zero energy.
Ignoring the minor effects of such correlation, one can approximate the
effective accumulation length $N$ in~\eqref{eq:processing-gain} for full-frame
Starlink processing as
\begin{equation}
  N \approx \Nsf (\Ns + \Ng) = 318912
\end{equation}
where $\Nsf = 302$ is the number of nonzero symbols in a frame, $\Ns = 1024$ is
the number of subcarriers, and $\Ng = 32$ is the CP length.  Thus, for the case
of full-frame correlation with completely known symbols, $L = N \approx 55$ dB.

\section{Data Corpus}
\label{sec:data-corpus}
The signal models and observations about data structures that follow are
supported by a large corpus of Starlink Ku-band captures spanning many
observation sessions and satellites, across which the reported properties were
observed consistently.  The corpus spans all generations of Starlink satellites
to date (blocks 1.0, 1.5, and 2.0-mini) and covers captures from October 2024
through June 2025.  Details about each capture are given in the Supplementary
Material.  To encourage follow-on work and ensure reproducibility, the
Supplementary Material contains decoded frame data for a high-quality subset of
1009 decoded frames that we call the \emph{exemplar frames}, which are from a
single capture of high-SNR data from one Starlink satellite.

\section{Signal Model}
\label{sec:signal-model}
To facilitate understanding of the OFDM signal processing techniques presented
in the sequel, this section presents a joint time- and frequency-domain signal
model for the Starlink Ku-Band downlink.  The model assumes the basic frame
structure detailed in \cite{humphreys2023starlinksignalstructure}.  In brief,
Starlink Ku-band downlink data are packaged into frames transmitted at a maximum
rate of \qty{750}{\hertz}.  Each frame consists of $\Nsf = 302$ OFDM symbol
slots of nominal length $\Tsym = \qty{4.4}{\micro\second}$ followed by an
unoccupied frame guard interval $\Tfg = \Tsym + \Tg$, where
$\Tg = 2/15~\unit{\micro\second}$ is the CP interval.  A sequence of frames from
a given Starlink satellite vehicle (SV) to a given service cell is conveyed by a
downlink beam whose SV-to-cell assignment remains fixed over a 15-second
fixed assignment interval (FAI) \cite{qin2025starlinktimingproperties}.

\subsection{Clock Models}
\label{sec:clock-models}
\begin{figure}[t]
  \centering
  \includegraphics[width=\columnwidth]{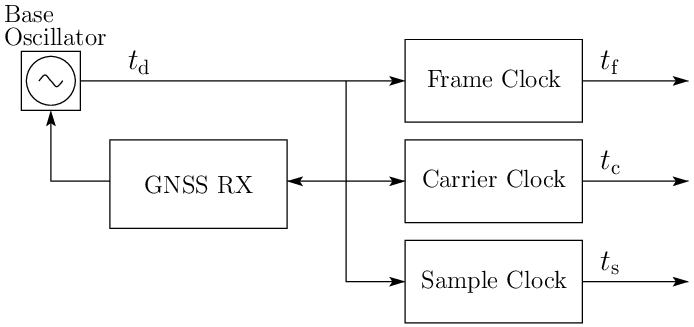}
  \caption{Cascading clock model for Starlink transmitter showing the
    relationship between base time $\td$, frame clock time $\tf$, carrier clock time
    $\tc$, and sample clock time $\ts$.}
  \label{fig:clock-model}
\end{figure}

Unlike GNSS and other DSSS-based waveforms, Starlink's symbols, frames, and
underlying carrier are not driven directly by the same clock---at least not for
block 1.0, 1.5, and 2.0 SVs \cite{qin2025starlinktimingproperties}. Accordingly,
we adopt the cascading clock model shown schematically in
Fig.~\ref{fig:clock-model}.  The GNSS-disciplined base time $\td$ drives the
downstream frame, carrier, and sample clocks.  These in turn drive signal
generation onboard Starlink SVs. The frame clock sets the start time of frames
within an FAI.  The carrier clock drives generation of the carrier signal used
to mix the baseband OFDM waveform to radio frequency (RF). The sample clock
drives the digital-to-analog converter used to reconstruct the digital waveform
as an analog baseband waveform.

Time according to these three clocks is related to true time $t$, or time
according to an ideal clock, by
\begin{align}
  \label{eq:1dtf}
  t &= \tf(t) - \dtf(t) \\
  \label{eq:1dtc}
  t &= \tc(t) - \dtc(t) \\
    \label{eq:1dts}
  t &= \ts(t) - \dts(t) 
\end{align}
where $\dtf(t)$, $\dtc(t)$, and $\dts(t)$ are the frame, carrier, and sample
clock offsets, respectively.  Given that this paper addresses the signal
processing of individual frames rather than the timing behavior of frame
sequences as in \cite{qin2025starlinktimingproperties}, the frame clock is not
discussed in further detail.

The time derivatives of $\dtc(t)$ and $\dts(t)$ are referred to as the carrier
and sample clock drifts.  For the temperature-controlled crystal oscillators
(TCXOs) apparently used in Starlink SVs \cite{qin2025starlinktimingproperties},
these may be modeled as constant over a frame interval
$\Tf = 1/750~\unit{\second}$.  This gives rise to a linear clock offset model
that is valid for $t_0 \leq t \leq t_0 + \Tf$:
\begin{align}
  \dts(t) &= \dtsn+\dtsd(t-t_0) \label{eq:1dmodel-sample} \\
  \dtc(t) &= \dtcn+\dtcd(t-t_0) \label{eq:1dmodel-carrier}
\end{align}
Here, $\dtcn$ and $\dtsn$ are the carrier and sample clock offsets at time
$t_0$, while $\dtcd$ and $\dtsd$ are the respective clock drifts, modeled as
piecewise constant from frame to frame. Substituting (\ref{eq:1dmodel-carrier})
into \eqref{eq:1dtc} and solving for $\tc(t)$ yields
\begin{align}
  \tc(t) &= t + \dtcn+\dtcd(t-t_0) \nonumber \\
  \label{eq:tcsolve}
         &= (1+\dtcd) t + (\dtcn- \dtcd t_0)
\end{align}
Similarly, $\ts(t)$ becomes
\begin{align}
    \label{eq:tssolve}
  \ts(t) &= (1+\dtsd) t + (\dtsn- \dtsd t_0)
\end{align}

\subsection{Transmitted Signal Model}
As mentioned, each Starlink frame contains $\Nsf = 302$ occupied OFDM symbol
intervals of length $\Tsym$.  We refer to the time-domain signal that applies
over the $i$th symbol interval of the $m$th frame as $x_{mi}(t)$, where
$i \in \Ids \define \{0, 1, \dots, \Nsf-1 \}$.  The first symbol interval
is occupied by $x_{m0}(t)$, a known time-domain sequence referred to as the PSS;
the second is occupied by $x_{m1}(t)$, a known time-domain OFDM symbol referred
to as the SSS.  The remaining $\{x_{mi}(t) \mid i \in [2, \Nsf -1]\}$ are
time-domain OFDM symbols containing a mix of header information, pilots, user
data, and---as will be shown---low-entropy codes.

Each $x_{mi}(t)$ for $i \in [1, \Nsf- 1]$ consists of $\Ns=1024$ mutually
orthogonal data subcarriers spread across the channel bandwidth
$\Fs=\qty{240}{\mega\hertz}$, resulting in a carrier spacing of
$\F=\Fs/\Ns$. Let $k \in \K \define \{0,1, \dots, \Ns-1 \}$ represent a single
subcarrier's assigned index. We adopt the typical OFDM subcarrier
index-to-frequency offset mapping $d[k]$, defined as
\begin{equation}
  \label{eq:dkdef}
  d[k] = \begin{cases} k, & k\in [0,\frac{\Ns}{2}-1] \\
    k - \Ns, & k \in [\frac{\Ns}{2},\Ns-1]
  \end{cases}
\end{equation}
               
Each subcarrier is modulated by a complex-valued information symbol
$X_{mik} \in \CC{mi}$, which encodes one or more bits of information depending
on the modulation scheme corresponding to the constellation $\CC{mi}$ (e.g.,
BPSK, 4QAM, etc., or a mixture of these).  Four mid-channel subcarriers, those
with indices in $\Kg = \{ 0, 1, 1022, 1023 \}$, form a gutter in which
$X_{mik} = 0$ for $k \in \K_\text{g}$.  Let $\Ido \define \Ids \setminus 0$ be
the symbol index set excluding the PSS, which is not a standard OFDM symbol.
Then for $i \in \Ido$ the baseband time-domain signal over $0 \leq t < \Tsym$
can be written as
\begin{equation}
  \label{eq:xmiDef}
  x_{mi}(t) = \frac{1}{\sqrt{\Ns}}\sum_{k\in \K} X_{mik} \exp \left(j2\pi d[k] \F (t-\Tg) \right)
\end{equation}
where $\Tg = \Ng / \Fs$ is the CP interval, $\Tsym = T + \Tg$ is the full symbol
interval, and $T = \Ns/\Fs$ is the useful (non-cyclic) symbol interval.  Note
that despite the introduction of $d[k]$ in \eqref{eq:xmiDef}, our definition of
$X_{mik}$ is identical to that of \cite{humphreys2023starlinksignalstructure}
for sampling at $t = 0, 1/\Fs, \dots, \Tsym \Fs$.  Let the OFDM symbol support
function be defined as
\begin{equation*}
  g_\text{s}(t) = \begin{cases} 2t/\Tg, & 0 \leq t < \Tg/2 \\
    1, & \Tg/2 \leq t < T + \Tg/2 \\
    1 - 2(t-T - \Tg/2)/\Tg, & T + \Tg/2 \leq t < \Tsym \\
    0, & \text{otherwise} \end{cases}
\end{equation*}
Then, assuming a start time $t_0 = 0$, the baseband time-domain signal over the
$m$th frame may be written as
\begin{equation}
  \label{eq:frameBasebandModel}
  x_m(t) =\sum_{i \in \Ids} x_{mi}(t - i \Tsym)g_\text{s}(t - i
  \Tsym), \quad 0 \leq t < \Tf
\end{equation}

Starlink OFDM signals are transmitted over one of eight channels spanning the
\qtyrange{10.7}{12.7}{\giga\hertz} band
\cite{humphreys2023starlinksignalstructure}.  In what follows, $\Fc{}$
represents an arbitrary channel center frequency selected from among the eight.
Assume $x_m(t)$ undergoes digital-to-analog conversion driven by $\ts(t)$, and
mixing from baseband to center frequency $\Fc{}$ driven by $\tc(t)$.  Then,
assuming $t_0 = 0$, we arrive at the transmitted $m$th-frame RF signal model
\begin{align}
  \tilde{x}_m(t) & = x_m\!\left[\ts(t)\right] \exp\left[j2\pi \Fc{} \tc(t) \right]
                   \nonumber \\
  \label{eq:frameRFModel}
                 &=  x_m\!\left[(1+\dtsd)t + \dtsn \right] \nonumber \\
                 & \qquad \times \exp\!\left\{j2\pi \Fc{} \!\left[ (1+\dtcd) t + \dtcn \right]
    \right\}
\end{align}

\subsection{Received Signal Model}
This section develops a discrete-time received signal model for the LEO-to-Earth
channel that includes sample frequency offset (SFO) and carrier
frequency offset (CFO) parameters.

Let $\vlos$ be the scalar velocity between the SV and receiver along the
line-of-sight, defined such that $\vlos < 0$ when the SV is moving towards the
receiver. Define the Doppler parameter $\beta \define \vlos / c$, where $c$ is
the free-space speed of light.  For signals from a LEO SV traveling at an
orbital altitude of \qty{550}{\kilo\meter} to a stationary ground receiver,
$\lvert\beta\rvert$ is bounded below $\qty{25}{ppm}$.  In practice, Starlink SVs
typically do not transmit below an elevation of $\sim\!\qty{40}{\degree}$, which
limits $\lvert\beta\rvert$ to below $\sim\!\qty{15}{ppm}$.  As the frame
interval $\Tf \approx \qty{1.33}{\milli\second}$ is short compared to the
timescale over which the SV-receiver geometry evolves, $\beta$ is modeled as
constant over each frame.

Let $\tau_\text{los}$ be the line-of-sight propagation delay for the leading
edge of the frame and $w(t)$ be receiver noise, modeled as complex-valued
zero-mean additive white Gaussian noise (AWGN) with two-sided power spectral
density $N_0/2$ per dimension and uncorrelated with the transmitted signal.
Assume that $\beta,\dtsd$, and $\dtcd$ are small enough that their second-order
products are negligible.  Then the analog RF received signal may be modeled as
\begin{align}
  \tilde{y}_m(t) &= \tilde{x}_m\left[(1-\beta)(t-\tau_\text{los})\right] + w(t) \nonumber \\
         &= x_m\!\left[(1+\dtsd-\beta) (t-\tau_\text{los}) + \dtsn \right] \nonumber \\
         &\quad \times \exp\!\left\{j2\pi (1+\dtcd-\beta)\Fc{} (t-\tau_\text{los}) + j2\pi\Fc{}\dtcn \right\} \nonumber \\
         &\quad + w(t) \label{eq:recv-signal}
\end{align}

Let the effects of $\tau_\text{los}$, $\dtsd$, and $\dtsn$ be combined into an
equivalent timing offset $\tau_{m}$, and those of $\tau_\text{los}$, $\dtcd$,
$\dtcn$ be combined into an equivalent phase offset $\phi_{m}$:
\begin{align*}
    \tau_{m} &= \tau_\text{los} - \frac{\dtsn}{1+\dtsd-\beta} \\
    \phi_{m} &=2\pi\Fc{}\dtcn - 2\pi(1+\dtcd-\beta)\Fc{}\,\tau_\text{los}
\end{align*}
Then the received signal model simplifies to
\begin{align*}
  \tilde{y}_m(t) &= x_m\!\left[(1+\dtsd-\beta) (t - \tau_{m}) \right] \\
  	     &\qquad \times \exp\!\left\{j\!\left[2\pi(1+\dtcd-\beta)\Fc{}  t + \phi_{m} \right] \right\} + w(t)
\end{align*}

For the data processing described in the sequel, signals were captured using a
highly directional antenna and sampled at a complex sampling rate of
\qty{250}{\mega sps} after mixing to baseband, then resampled at $\Fs$ such that
all subsequent operations may proceed at the native Starlink information symbol
rate.  The receiver clock driving our capture equipment was a GNSS-disciplined
OCXO whose time can be taken as synonymous with true time $t$.  To reduce
complexity, we aggregate the clock drift and Doppler parameters into an
effective SFO factor $\betas \define -\dtsd+\beta$ and an effective CFO factor
$\betac \define -\dtcd+\beta$.  Let $t_m$ be the true start time of the $m$th
frame as received, which is governed by Doppler and by the satellite frame
clock.  Then, for a sampling interval $\Ts = 1/\Fs$ and receiver sample index
$n \in \mathbb{Z}$ on $t_m \leq n\Ts < t_m + \Tf$, we arrive at the baseband
discrete-time signal model
\begin{equation}
  \begin{split}
    \tilde{y}_m[n] &= x_m\!\left[(1-\betas)\,(n - n_{m})\Ts\right] \\ 
                   &\qquad \times \exp\!\left(-j2\pi \, \betac \Fc{} \, n\Ts +
                     j\phi_{m} \right) + w(n\Ts)
  \end{split}
\end{equation}
where $n_{m} \in \mathbb{R}$ satisfies $n_{m} \Ts = \tau_{m}$.  A continuous sequence
of received samples across both occupied and unoccupied frame slots is given by
\begin{equation}
  \label{eq:continuousSequenceyn}
  \tilde{y}[n] = \sum_{m \in \mathbb{Z}} \tilde{y}_m[n - t_m/\Ts]g_\text{f}(n\Ts - t_m), \quad n \in \mathbb{Z}
\end{equation}
where
\begin{equation*}
  \label{eq:frameSupportFcn}
  g_\text{f}(t) = \begin{cases} 1, & 0\leq t < \Tf \\
0,  & \text{otherwise}\end{cases}
\end{equation*}
is the frame support function.  Note that, as regards the sample index and frame
support function limits, the wide frame guard interval $\Tfg$---during which no
signal is present---allows us to ignore small deviations from $\Tf$ in the
received signal's frame period.

\section{Demodulation}
\label{sec:demodulation}
This section describes the demodulation procedure used to correct residual
synchronization errors and recover the transmitted information symbols from a
single received frame.  Here, we follow the indexing convention established in
Section~\ref{sec:signal-model}: the received, equalized, phase-aligned, and
hard-decoded counterparts to $X_{mik}$ are denoted $\bar{Y}_{mik}$,
$\tilde{Y}_{mik}$, $Y_{mik}$, and $\hat{X}_{mik}$, respectively, with
corresponding OFDM-symbol-level vectors $\bYmi$, $\tYmi$, $\vbY{mi}$,
$\hXmi \in \mathbb{C}^{|\K|}$.  

In brief overview, the demodulation process proceeds as follows.  Frame $m$ is
detected and coarsely acquired by matched filtering against a Doppler-adjusted
PSS+SSS local replica, whereupon standard OFDM processing yields, for each
symbol index $i\in \Ids$, a noisy received vector $\bYmi$.  Residual SFO and CFO
errors are then estimated and compensated.  Finally, hard-decoding decisions are
made to produce the recovered information symbol vectors $\hXmi$.

\subsection{Coarse Frame Acquisition}
Individual Starlink frame detection and acquisition is accomplished through
matched filtering with a Doppler-adjusted local baseband replica. Define the
coherent concatenation of the time-domain PSS and SSS functions from
\cite{humphreys2023starlinksignalstructure} as
\begin{equation*}
\label{eq:coherentCombPssSss}
  x_{m01}(t) \define \begin{cases}
  x_{m0}(t), & 0 \leq t < \Tsym \\
  x_{m1}(t - \Tsym), & \Tsym \leq t < 2\Tsym \\
  0, & \mbox{otherwise}
\end{cases}
\end{equation*}
Because the PSS and SSS are identical for all $m \in \mathbb{Z}$ and all SVs,
$x_{m01}(t)$ can be used for signal detection on all captured data. The full
discrete-time local replica used for correlation is the sampled product of the
Doppler-adjusted $\xlr(t)$ and a complex exponential:
\begin{equation*}
  \tylr[n,\beta] \define \xlr\left[n\Ts(1-\beta)\right] \exp\left(-j2\pi\beta
    \Fc{} n\Ts\right)
\end{equation*}
Sampling of the local replica is limited to $n \in [0,M(\beta)-1]$ with
$M(\beta) = \left\lceil 2\Tsym/(1-\beta)\Ts \right\rceil$.  For $\tilde{y}[n]$
given as in \eqref{eq:continuousSequenceyn}, the complex ambiguity function is
formed as
\begin{gather}
  \label{eq:Ratellbeta} 
  R[k,\beta] = \!\!\sum_{n=0}^{M(\beta)-1} \tilde{y}[n+k] \, \ylr^*[n,\beta]
\end{gather}
The nearest-sample timing offset estimate $\hat{n}_m$ and initial Doppler
estimate $\hat{\beta}_m$ are obtained via a two-dimensional search over lag and
Doppler hypotheses:
\begin{equation*}
  (\hat{n}_m, \hat{\beta}_m) = \argmax_{k, \beta} \left| R[k, \beta] \right|
\end{equation*}
These estimates are accepted if $| R[\hat{n}_m, \hat{\beta}_m]|$
exceeds a detection threshold, whereupon the $m$th Starlink frame is 
extracted from the continuous stream as
\begin{equation*}
  \tilde{y}_m[n] = \tilde{y}[n + \hat{n}_m], \quad n = 0, 1, \dots, \Nsf (\Ns + \Ng) - 1
\end{equation*}
Using the initial Doppler estimate $\hat{\beta}_m$, the extracted frame is
coarsely Doppler compensated via time resampling and carrier frequency
compensation. The resulting signal can be modeled in terms of residual
synchronization errors as
\begin{align}
  \label{eq:sigWithSynchErrors}
  \bar{y}_m[n] &= x_m \left[(1-\dbetas)\,(n-\delta n)\Ts\right] \\ 
  &\quad \times \exp\left[-j2\pi \, \dbetac \Fc{} \, n\Ts + j \phi_{m} \right] +
    w(n\Ts) \nonumber
\end{align}
where $\delta n = n_{m} - \hat{n}_{m}$ denotes the fractional residual timing
offset, $\dbetas = \betas-\hat{\beta}_m$ the residual SFO factor, and
$\dbetac = \betac-\hat{\beta}_m$ the residual CFO factor.  We assume
$\abs{\delta n} \Ts < \Tg$ and $\abs{\dbetac} \Fc{} \ll \F$ so that CP removal
and frequency transformation yield negligible inter-symbol interference (ISI) and
inter-carrier interference (ICI).  Next, $\tilde{y}_m[n]$ is segmented into $\Nsf$ blocks
of $\Ns + \Ng$ samples each.  Each block undergoes CP removal and conversion to
the frequency domain via the FFT.  In the frequency domain, the $i$th OFDM
symbol's $k$th received data symbol $\bar{Y}_{mik}$ can be modeled as the following
function of the transmitted data symbol $X_{mik}$:
\begin{align}
  \label{eq:rxSymbolModel}
  \bar{Y}_{mik} &= \alpha_{mik} X_{mik} + V_{mik} \\
  \alpha_{mik} &= Z_{m} \, H_k \, \exp(-j 2 \pi d[k] \F \tau_{mi} + j \phi_{mi})
\end{align}
Here, $V_{mik}$ denotes frequency-domain AWGN, $H_k$ is the frequency-selective
channel response at subcarrier $k$ normalized such that $H_0 = 1$, and
$Z_{m} = \sqrt{g} \exp(j\theta_m)$ is a frame-specific complex constant,
independent of $k$, with channel gain $g$ and initial phase offset $\theta_m$.
The phase offset $\theta_m$ does not appear to be arbitrary from frame to frame,
but our efforts at modeling it have so far not successfully generalized to all
frames and all Starlink SVs.  The per-symbol time delay $\tau_{mi}$ and
phase offset $\phi_{mi}$ relate to $\dbetas$ and $\dbetac$ as follows:
\begin{align}
    \label{eq:tau_mi}
  \tau_{mi} &= \delta n\Ts + i \Tsym \left( \frac{\dbetas}{1-\dbetas} \right) \approx \delta n\Ts + i \Tsym \dbetas \\
  \label{eq:phi_mi}
  \phi_{mi} &= \phi_{m} - 2\pi i \Tsym \dbetac \Fc{}
\end{align}

\subsection{Channel Equalization}
\label{sec:channel-equalization}
The channel transfer function $H_k$ may not have a flat amplitude response or a
linear phase response over the entire
\qty[quantity-product={-}]{240}{\mega\hertz} band of a Starlink channel, in
which case channel equalization is helpful before proceeding with demodulation.
As a native time-domain signal, the PSS is excluded from this process; thus, our
concern is the symbol index set $\Ido \define \Ids \setminus 0$.

Let $\Kl \define \K \setminus \Kg$ denote the set of loaded (non-gutter)
subcarriers. The channel transfer function can be identified by using the known
transmitted SSS symbols $X_{m1k}$ for all $k\in \Kl$.  Suppose that the
corresponding received demodulated values $\bar{Y}_{m1k}$ have been determined
using the techniques from the preceding subsection, and suppose we define
$\phi_{m1} \define 0$.  Then the following relationship holds true for all $k\in \Kl$:
\begin{equation}
  \label{eq:preEqualizationWithExp}
  \bar{Y}_{m1k} = Z_m  H_k \exp(-j 2 \pi d[k] \F \tau_{m1})  X_{m1k} + V_{m1k}
\end{equation}
Under the constraint $H_0 = 1$, a pre-equalization maximum-likelihood estimate
of $\tau_{m1}$ can be obtained from $\bar{Y}_{m1k}X^*_{m1k}$, $k \in \Kl$ in the
same way that frequency is estimated in \cite{rife1974single}.  This allows the
exponential factor in \eqref{eq:preEqualizationWithExp} to be eliminated,
yielding
\begin{equation}
  \label{eq:preEqualizationWithoutExp}
  \bar{Y}_{m1k} = Z_mH_kX_{m1k} + V_{m1k}
\end{equation}
One can then estimate $H_k$ (to within the constant $Z_m$) by computing
$\breve{H}_k = \bar{Y}_{m1k}/X_{m1k}$ for all $k\in \Kl$.  To mitigate the
effects of noise, one may apply a filter (e.g., a Savitzky–Golay filter) or a
spline fit to the $\breve{H}_k$ for all $k\in \Kl$ to produce a smoothed
function $\tilde{H}_k$.  One obtains the value $\tilde{H}_0$ from the smoothed
function and uses this to compute a normalized smoothed estimate of the transfer
function, $\hat{H}_k = \tilde{H}_k/\tilde{H}_0$.  This technique can be used for
multiple successive frames of a given receiver/satellite/band combination to
refine the channel transfer function estimate.

\begin{figure}[t]
  \centering
  \includegraphics[trim=16 5 30 20,clip,width=\linewidth]{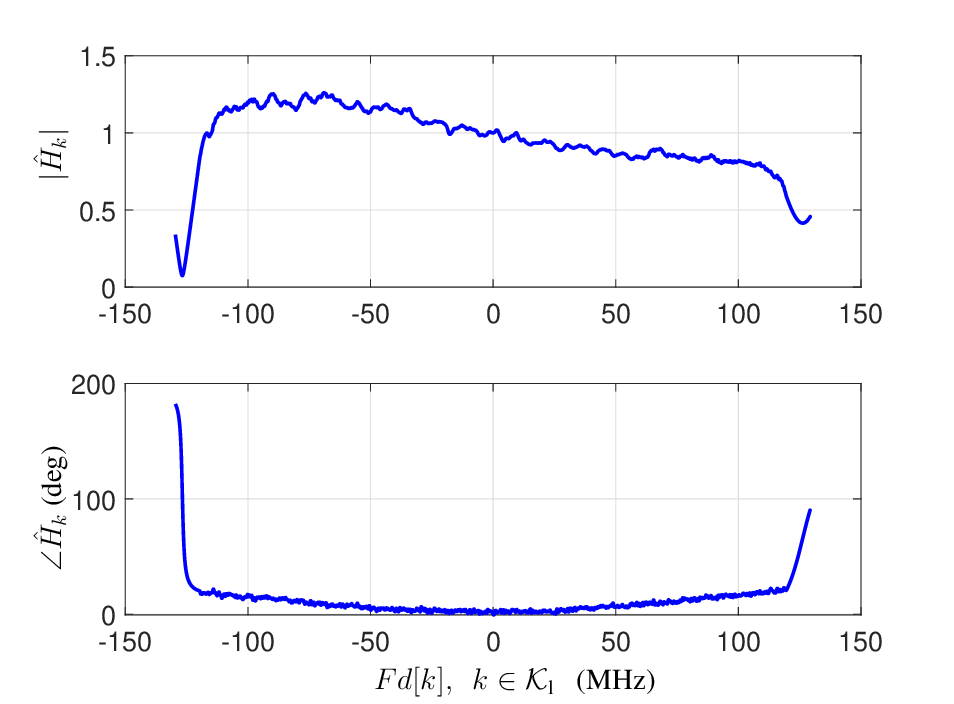}
  \hfill
  \caption{Example estimated Starlink channel transfer function $\hat{H}_k$ as a
    function of offset frequency $\F d[k]$ for $k\in \Kl$.  This transfer
    function is for a data recording system that uses an Ettus X410 USRP with
    native sampling at \qty{250}{\mega\hertz} and a
    \qty[quantity-product={-}]{1.2}{\meter} steered-dish antenna.}
  \label{fig:channel-tf}
\end{figure}

Fig. \ref{fig:channel-tf} shows an example estimated transfer function.  Note
how the amplitude falls off going from the low-frequency section to the
high-frequency section.  It also falls off rapidly near the two band edges.  The
phase stays relatively flat through most of the band, but changes rapidly near
the band edges at the same frequencies that experience sharp amplitude roll-off.

The frame-specific constant $Z_m$ is estimated as
$\hat{Z}_{m} = \sqrt{\hat{g}} \exp(j\hat{\theta}_m)$, where $\hat{g}$ is the
estimated channel gain and $\hat{\theta}_m$ is the estimated initial carrier
phase:
\begin{align*}
    \hat{g} = \frac{1}{|\Kl|}\sum_{k\in\Kl}|\bar{Y}_{m1k}/\hat{H}_k|^2,\quad
    \hat{\theta}_m = \angle \sum_{k\in\Kl} (\bar{Y}_{m1k}/\hat{H}_k) X_{m1k}^*
\end{align*}
Normalization by $\hat{H}_k$ and $\hat{Z}_m$ equalizes the received symbols for
all $k\in \Kl$ and for all $i\in \Ido$:
\begin{equation*}
 \tilde{Y}_{mik} = \bar{Y}_{mik}/(\hat{Z}_m \hat{H}_k) 
\end{equation*}
The equalized symbols can be modeled by the following simplification of
\eqref{eq:rxSymbolModel}
\begin{equation}
  \label{eq:rxSymbolModelTilde}
  \tilde{Y}_{mik} = \exp(-j 2 \pi d[k] \F \tau_{mi} +
  j \phi_{mi}) X_{mik} + \tilde{V}_{mik} 
 \end{equation}
 with $\tilde{V}_{mik}$ being complex AWGN and $\tau_{mi}$ and $\phi_{mi}$ given as
 in \eqref{eq:tau_mi} and \eqref{eq:phi_mi}.

\subsection{Residual Synchronization Parameter Estimation}
As is clear from \eqref{eq:rxSymbolModelTilde}, each $\tYmi$, $i \in \Ido$
remains distorted by residual synchronization errors and phase misalignment.  We
remedy this via the multi-stage process described
in~\cite{qin2025toadopplerplans} and summarized below.

First, each constellation $\CC{mi}$ must be identified for $i \in \Ido$.
Starlink employs several different quadrature amplitude modulation (QAM) schemes
to encode a sequence of digital data onto the $\Ns$ complex information symbols
that modulate the subcarriers of each OFDM symbol.  We distinguish 4QAM from
QPSK constellations according to the convention
\begin{equation*}
  \Cqpsk = \{1,j,-1,-j\}, \quad
  \Cfqam = \Cqpsk \exp(j\pi/4)
\end{equation*}
Thus, $\Cfqam$ is equivalent to $\Cqpsk$ modulo a \qty{45}{\degree} rotation.
We exclude the edge pilots (formally introduced in Section
\ref{sec:edge-pilots}) from the determination of $\CC{mi}$ because edge pilot
symbols are uniformly 4QAM.  Let $\Kp$ be the subcarrier indices corresponding
to the edge pilots, and let $\Klnp \define \Kl \setminus \Kp$ be the set of
loaded non-pilot subcarrier indices.  Only subcarriers $k \in \Klnp$ are
involved in constellation identification.

We identify $\CC{mi}$ by estimating its cardinality $\lvert\CC{mi}\rvert$ from
$\tYmi$ via $k$-means clustering.  If the estimated cardinality is 16 or 32 we
select $\hCC{mi} = \Cstqam$ or $\hCC{mi} = \Cttqam$ as appropriate; for
cardinality 4, we initially assume $\hCC{mi} = \Cqpsk$ and defer the QPSK/4QAM
disambiguation to after residual synchronization compensation.  For rare cases
of other estimated cardinality (e.g., 8), we declare the OFDM symbol to have
composite modulation within $\Klnp$ and discard it.  Let $\Idm \subseteq \Ido$
be the set of all OFDM symbols retained after this process as applied to the
$m$th frame.

Given $\hCC{mi}$, we estimate the residual delay and phase parameters
$\tau_{mi}$ and $\phi_{mi}$ from \eqref{eq:rxSymbolModelTilde} using a
symbol-marginalized maximum-likelihood (ML) formulation. Since channel
equalization has already normalized the symbols to approximately unit power, the
ML formulation focuses solely on phase and delay estimation.  Let
$\vb{\theta}_{mi} \define [\tau_{mi}, \phi_{mi}]$ and let
\begin{gather*}
   \Lambda_{\tYmi}(\vb{\theta}_{mi}\mid\hCC{mi})
  = \prod_{k\in\Kl} p(\tilde{Y}_{mik}\mid\vb{\theta}_{mi},\hCC{mi})
\end{gather*}
be the likelihood function of the vector measurement $\tYmi$, where
$p(\tilde{Y}_{mik}\mid\vb{\theta}_{mi},\hCC{mi})$ is the likelihood of $\tilde{Y}_{mik}$
for a given $\vb{\theta}_{mi}$ assuming $\CC{mi} = \hCC{mi}$.  For data-bearing information
symbols, the likelihood function is obtained by marginalizing over the possible
values for the unknown symbol $X_{mik}$, assuming complex Gaussian noise under a
uniform prior on $\hCC{mi}$:
\begin{align}
  \label{eq:margOverCmi}
  p(\tilde{Y}_{mik}\mid\vb{\theta}_{mi},\hCC{mi})
  = \frac{1}{|\hCC{mi}|} \sum_{c\in\hCC{mi}}
    p(\tilde{Y}_{mik}\mid X_{mik}=c,\vb{\theta}_{mi}) 
\end{align}

Pilot information symbols are handled as the special case where $X_{mik}$ is
known and no marginalization is required. The ML estimate of $\vb{\theta}_{mi}$
may be found by
\begin{equation}
  \label{eq:thetaMLhat}
  \hat{\vb{\theta}}_{mi}
  = \argmax_{\vb{\theta}_{mi}} \log \Lambda_{\tYmi}(\vb{\theta}_{mi}\mid\hCC{mi}) \\
\end{equation}

This process yields a sequence of per-OFDM-symbol delay and phase estimates
$\{\hat{\tau}_{mi}, \hat{\phi}_{mi} \mid i \in \Idm\}$.  While these estimates
are statistically efficient on a per-symbol basis, they are noisy and do not
obey the temporal structure imposed by \eqref{eq:tau_mi} and \eqref{eq:phi_mi}
wherein $\tau_{mi}$ and $\phi_{mi}$ vary linearly with OFDM symbol index $i$.
To exploit this structure and obtain a unified set of frame-level residual
synchronization parameters, the per-symbol ML estimates are combined by a joint
ML fit as follows.  Let $\vbh{\phi}_m = [\hat{\phi}_{mi} \mid i \in \Idm]$ and
$\vbh{\tau}_m = [\hat{\tau}_{mi} \mid i \in \Idm]$ respectively denote column
vectors of per-OFDM-symbol phase and delay estimates obtained from the preceding
ML stage for frame $m$. Likewise, let the vector $\vb{i} = [i \mid i \in \Idm]$
contain the relevant symbol indices, and $\vb{1}$ be the all-ones vector of
matching dimension.  Let
$\vb{\psi}_m \define [\phi_{m0}, \dbetac, \tau_{m0}, \dbetas]$ contain the
parameters to be estimated, and define $\vb{J}(\vb{\psi}_m)$ as
\begin{equation*}
  \vb{J}(\vb{\psi}_m) \define
\left[  \ba{lcr}
    \vbh{\phi}_m &-&
    \big(\phi_{m0} \vb{1} - 2\pi \Tsym \Fc{} \dbetac\, \vb{i}\big) \\
    \vbh{\tau}_m &-&
    \big(\tau_{m0} \vb{1} - \Tsym\, \dbetas\, \vb{i}\big)
\ea\right]
\end{equation*}
Minimizing the squared magnitude of this function yields the ML estimate for
$\vb{\psi}_m$:
\begin{equation}
  \label{eq:mlEstPsi}
  \hat{\vb{\psi}}_m
  = \argmin_{\vb{\psi}_m}
    \big\| \vb{J}(\vb{\psi}_m) \big\|_2^2
\end{equation}
Compensation of the residual synchronization errors yields the phase-aligned
symbols for all $k \in \Kl$ and $i \in \Idm$:
\begin{equation}
  \label{eq:compensatedYmik}
  Y_{mik} = \tilde{Y}_{mik} \exp\!\left(-j\hat{\phi}_{mi} + j 2\pi d[k] \F \hat{\tau}_{mi}\right)
\end{equation}

\begin{figure}[t]
  \centering
  \includegraphics[trim=0 0 7 0,clip,width=\columnwidth]{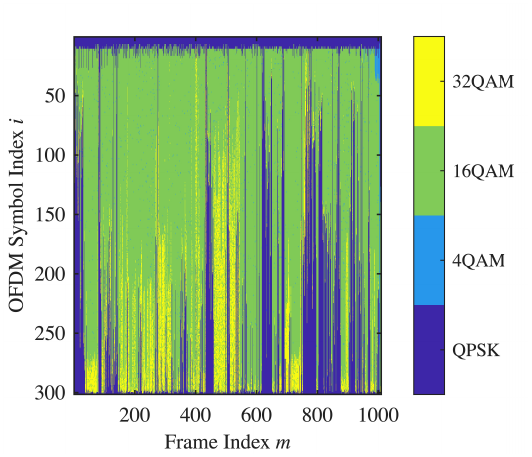}
  \caption{Modulation scheme estimates for the set of 1009 exemplar frames, with
    each frame comprising $\vert \Ido \vert = 301$ OFDM symbols.  Frames and
    symbols are ordered along the horizontal and vertical axes, respectively.
    The color of each cell indicates the estimated $\CC{mi}$ type for
    $m \in \{1,\dots,1009\}$ and $i \in \Ido$. Note how the first few symbols
    (starting with the SSS) of every frame are QPSK-modulated. These symbols
    constitute the variable-length frame header.}
  \label{fig:modulation_map}
\end{figure}
\subsection{Information Symbol Detection}
With the phase-aligned information symbol vector $\vbY{mi}$ available for all
$i \in \Idm$, the QPSK/4QAM ambiguity for cardinality-4 symbols is resolved: For
each $\vbY{mi}$, we select $\hCC{mi} = \Cqpsk$ or $\hCC{mi} = \Cfqam$ depending
on the relative alignment against $\vbY{m(i-1)}$, with $\vbY{m1}$, the
phase-aligned SSS, providing an absolute reference.  

Each hard-decoded symbol estimate $\hat{X}_{mik}$ is then obtained by mapping
$Y_{mik}$ to its nearest constellation point in $\hCC{mi}$:
\begin{equation*}
  \hat{X}_{mik} = \argmin_{c \in \hCC{mi}} \, \vert Y_{mik} - c \vert
\end{equation*}
This process produces the frame-level information symbol matrix
\begin{equation*}
  \hXm
  \define \left[
  \hat{\X}_{mi} \mid i \in \Idm \right] \in \mathbb{C}^{|\K|\times|\Idm|}
\end{equation*}
whose columns correspond to hard-decoded frequency-domain OFDM symbols.
Fig.~\ref{fig:modulation_map} illustrates the estimated modulation scheme for
each OFDM symbol across the exemplar frames.

\subsection{Alternate Residual Synchronization Parameter Estimation}
A more nearly optimal method was developed to remove the residual carrier
Doppler shift error $\dbetac$ and phase $\phi_{m0}$ when determining the edge
pilot symbols.  In addition to marginalizing over $\CC{mi}$, as in
\eqref{eq:margOverCmi}, this method marginalized the probability distribution
for $\tYmi$ over the 4 possibilities for the $i$th symbol interval, namely QPSK,
4QAM, 16QAM, and 32QAM.  This marginalization procedure assumed that each of
these 4 possibilities is equally likely, despite evidence to the contrary.  The
marginalized probability density function for $\tYmi$ was conditioned on the
residual phase angles of its $\tilde{Y}_{mik}$ components
\begin{equation*}
   \phi_{mik} = \phi_{m0} - 2 \pi (\dbetac \Fc + \dbetas F d[k]) (\Tsym/[1 - \hat{\beta}_m]) i
\end{equation*}

The negative-log-likelihood functions for all of the
$\tilde{\boldsymbol{Y}}_{mi}$ symbol vectors were summed for $i \in \Idt$, where
$\Idt \define \{2, 3, \dots, 301\}$ denotes OFDM symbol indices excluding the
PSS ($i=0$) and SSS ($i=1$).  Two additional terms were added to penalize
non-zero $\dbetac$ and the difference between $\phi_{m0}$ and its estimate based
purely on the PSS and SSS symbols.  The resulting negative-log-likelihood cost
function was minimized to obtain estimates of $\dbetac$ and $\phi_{m0}$.  This
minimization applied the approximation $\dbetas = \dbetac$ in the $\phi_{mik}$
formula above.  It started with a brute-force search over an extended $\dbetac$
range in order to get near the global minimum instead of a local minimum.
Afterwards, it applied Newton's method to find the optimal estimates of
$\dbetac$ and $\phi_{m0}$.  These estimates were then used to compute the
phase-aligned, demodulated symbols $\{Y_{mik} \mid  i \in \Idt, k \in \Kl \}$.

All loaded subcarriers---including those for the edge pilots---were included in
this data fitting procedure so that the resulting $Y_{mik}$ values could be used
generally; e.g., to determine whether or not there exist edge pilots and other
predictable symbols.

\section{Edge Pilots}
\label{sec:edge-pilots}
We developed a tailored averaging procedure to search for information symbols
that repeat exactly from frame to frame in the Starlink downlink.  The procedure
computed $|\Idt| \times |\Kl| = 306000$ averages of the received, equalized, and
phase-aligned symbols $Y_{mik}$---one average per unique combination of
$i \in \Idt$ and $k \in \Kl$.  The averaging was performed across frames (i.e.,
over multiple $m$ values) from multiple time intervals, multiple Starlink
frequency bands, and multiple Starlink satellites (see the Supplementary
Material for details).  If the same coefficient $X_{mik}$ is present for all
frames, then the resulting average approaches $X_{mik}$.  If, instead, $X_{mik}$
varies with $m$, then the average approaches zero.  Combinations $(i,k)$ that
produced large absolute values of the corresponding averages were examined to
ensure that the individual $Y_{mik}$ values used to produce the given average
appeared to be noisy versions of the same frame-to-frame-repeated symbol
$X_{ik}$.

Application of this averaging procedure revealed two bands of 4QAM-modulated
pilot symbols, one near each edge of a Starlink channel's allocated bandwidth.
The existence of these edge pilots was reported in
\cite{mccormick2024ofdm,kozhaya2025unveiling}, but not their symbol values,
which are published here.  Each of the two edge pilot bands occupies 8
subcarriers.  The relevant indices of both bands are
$\Kp \define \{488, 489, ..., 495, 528, 529, ..., 535\}$.  A 4QAM pilot
coefficient is modulated onto each $k\in \Kp$ for each $i \in \Idt$.

Notably, we found that the edge pilots appear to be identical, not only across
all frames from the same Starlink SV, but also across all SVs in the
constellation.  The pilot coefficient $X_{mik}$ takes the following form for
OFDM symbol $i \in \Idt$ and subcarrier $k \in \Kp$:
\begin{align}
\label{eq:pilotCoeff}
  X_{mik} &= \exp\!\left[j \frac{\pi}{2}\left(\spik + \frac{1}{2}\right)\right] \\
\label{eq:spik}
  \spik &= \left\lfloor  
          \frac{\qpk}{4^{(301-i)}}
          \right\rfloor  \bmod 4
\end{align}
The quantity $\spik \in \{0,1,2,3\}$ for $i \in \Idt$ and $k \in \Kp$ is a
function of $\qpk$, a 150-digit hexadecimal number given in
Appendix~\ref{app:pilot-codes} for each $k \in \Kp$.  The foregoing results are
identical to those in \cite{m_psiaki2025iongnss_leopanel} except for slight
notational adjustments and for the basis of representation (base-4
vs. hexadecimal).

\section{Low-Entropy QPSK Symbols}
\label{sec:low-entropy-qpsk}
This section reveals the surprising fact that nearly all QPSK-modulated
information symbols in the Starlink Ku-band downlink are not high-entropy user
data.  Rather, outside a short header region, these symbols exhibit a highly
repetitive structure across the full capture corpus: deviations relative to a
reference set of information symbols called the \emph{reference template} are
BPSK-valued and follow a simple periodicity with deterministic shifts across
OFDM symbols.  This structural repetition enables construction of long, near
noise-free local replicas for matched filtering, boosting processing gain.

\subsection{Reference Template}
\label{sec:reference-template}
Our analysis of low-entropy structure (aside from the edge pilots) in decoded
frame data is based on the exemplar frames mentioned in Section
\ref{sec:data-corpus}, with discovered patterns verified against data from the
larger corpus.  The exemplar frames, denoted $\{\hXm \mid m \in \Mex\}$ with
$\Mex = \{1, 2, \dots, 1009\}$, all come from a single capture of data from one
SV.  They are ordered chronologically, with no frames from occupied slots
discarded.  The pre-correlation SNR for each exemplar frame is greater than
\qty{9}{\decibel} and the average over the exemplar set is \qty{13.8}{\decibel}.

To identify portions of the Starlink frame that repeat from frame to frame, we
compute pairwise cross-correlations. We begin by defining several useful sets
and measures.  Let the set of non-SSS QPSK-modulated symbol indices for frame
$m$ be
\begin{equation*}
  \IQm \define \{\, i \in \Idt \mid \hCC{mi} = \Cqpsk \,\}
\end{equation*}
and the corresponding QPSK ratio be $\rho_m \define {|\IQm|}/{|\Idt|}$.  For a
pair of frames $\{m,n\}$, define the indices of symbols that are QPSK in
both frames as
\begin{equation*}
  \IQmn \define
  \{\, i \in \Idt \mid
      \hCC{mi} = \hCC{ni} = \Cqpsk \,\}
\end{equation*}
The corresponding shared QPSK ratio is $\rho_{mn} \define {|\IQmn|}/{|\Idt|}$.
Define the cross-correlation between frames $m$ and $n$, restricted to their
shared QPSK OFDM symbols and loaded non-pilot subcarriers, as
\begin{equation}
  \label{eq:Rmn}
  R_{mn} \define
  \sum_{i \in \IQmn} \sum_{k \in \Klnp}
  \hat{X}_{mik}^* \hat{X}_{nik}
\end{equation}

\begin{figure}[t]
  \centering
  \includegraphics[width=\columnwidth]{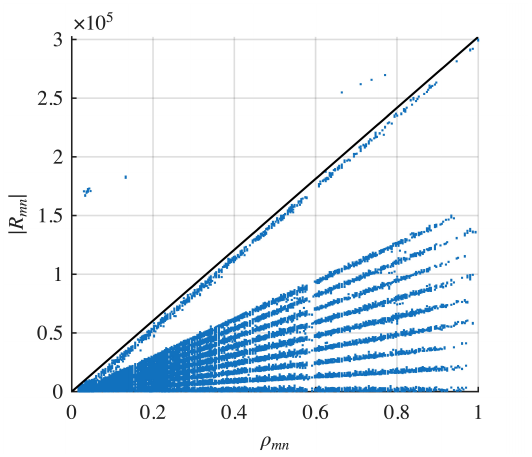}
  \caption{$\lvert R_{mn} \rvert$ vs.\ $\rho_{mn}$ for $m,n \in \Mex$, with
    $n<m$ to avoid duplicate pairwise comparisons. The black line shows the
    upper limit of $\lvert R_{mn} \rvert$, which obtains when all QPSK-modulated
    OFDM symbols carry precisely the same information. A small number of points
    lie above this theoretical limit. Investigation revealed that these
    correspond to consecutive frames (i.e., $\{m,m+1\}$) where the SV appears to
    have retransmitted duplicate data.}
  \label{fig:qpsk-ratio-corr}
\end{figure}

If QPSK-modulated OFDM symbols carried high-entropy user data from frame to
frame, then $\lvert R_{mn} \rvert$ would concentrate near zero regardless of
$\rho_{mn}$. An empirical analysis of $\lvert R_{mn} \rvert$ instead shows a
clear dependence on $\rho_{mn}$, as illustrated in
Fig.~\ref{fig:qpsk-ratio-corr}: the data form a structured fan with vertically
quantized bands. This behavior indicates that QPSK-modulated OFDM symbols
exhibit various degrees of correlation across frames and therefore warrant
focused analysis.

Further study revealed that pure-QPSK frames, those for which $\rho_m = 1$,
exhibit striking consistency: Nearly all information symbols agree across
pure-QPSK frames in the exemplar set.  Exceptions are a mix of rare hard-decoding
errors and some genuine frame-to-frame variation within the frame header. Let
the indices of pure-QPSK frames be denoted
$\Mexp \define \{ m \in \Mex \mid \rho_m = 1 \}$.  We define the
\textit{reference template} $\Tref \in \mathbb{C}^{|\Klnp|\times|\Idt|}$ as the
element-wise mode of the hard-decoded symbols $\{\hXm \mid m \in \Mexp\}$.  In
other words, the $(i,k)$th element of $\Tref$ for $i \in \Idt$ and $k \in \Klnp$
is defined as
\begin{equation}
  T_{ik} \define \argmax_{c \in \Cqpsk} \big| \{\, m \in \Mexp \mid \hat{X}_{mik} = c \,\} \big|
\end{equation}
We provide the full reference template $\Tref$ in the Supplementary Material.


\subsection{Tessellation Codes}
\label{sec:t-codes}
This section leverages the reference template $\Tref$ to expose the highly
regular structure of QPSK-modulated OFDM symbols, then introduces the repeating
bit patterns that underlie this structure.

To quantify deviations from the reference template, we restrict
$\{\hXm \mid m \in \Mex\}$ to QPSK-modulated symbols, then compute the
element-wise product between the decoded information symbol and the conjugate of
the corresponding reference template element. Let $D_{mik}$ denote the $(i,k)$th
element of the deviation, with corresponding OFDM-symbol-level vector
$\D_{mi} \in \mathbb{C}^{|\Klnp|}$ and frame-level matrix
$\D_m \in \mathbb{C}^{|\Klnp|\times|\IQm|}$. The element is defined for
$m \in \Mex$, $i \in \IQm$, and $k \in \Klnp$ as
\begin{gather}
  D_{mik} \define \hat{X}_{mik} T_{ik}^{*}
\end{gather}
Inspection of $\D_m$ reveals regular patterns, as follows.
\paragraph*{BPSK Deviations}
A histogram analysis of $\Tref$, and of the set of all QPSK-valued OFDM symbols
in the exemplar frames $\{\hXmi \mid m \in \Mex, i \in \IQm\}$, and of the set
of all deviation matrices $\{\D_m \mid m \in \Mex\}$ revealed that, whereas the
former two are uniformly distributed on $\Cqpsk$, the deviation matrices are in
fact BPSK-modulated; in particular, $D_{mik} \in \{1,-1 \}$.  Thus,
QPSK-modulated OFDM symbols can be expressed as the product of a BPSK-valued
sequence and the corresponding column of $\Tref$.  Analysis of the sequence of
deviation vectors $\{\D_{mi} \mid m \in \Mex, i \in \IQm\}$ reveals that the
QPSK-modulated portion of each frame comprises two distinct temporal regions: a
header region and a T-code region.

\begin{figure}[t]
  \centering
  \includegraphics[width=\columnwidth]{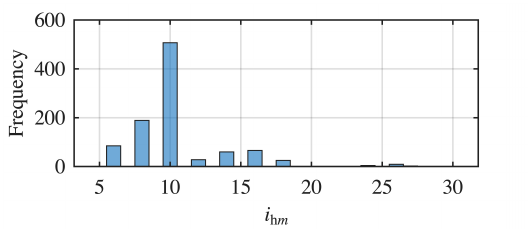}
  \caption{Distribution of the header length $\ihm$ across the exemplar
    frames.}
  \label{fig:mhead-hist}
\end{figure}
\paragraph*{Header}
After the PSS and SSS, each frame begins with a sequence of QPSK-modulated OFDM
symbols that form a header of variable length whose information content differs
from the highly regular structure observed later in the frame. In this paper we
do not attempt to interpret the header's content semantically; however,
preliminary analysis shows multi-modal frame-to-frame repetition.

\begin{figure}[t]
  \centering
  \includegraphics[trim=4 0 10 0,clip,width=\columnwidth]{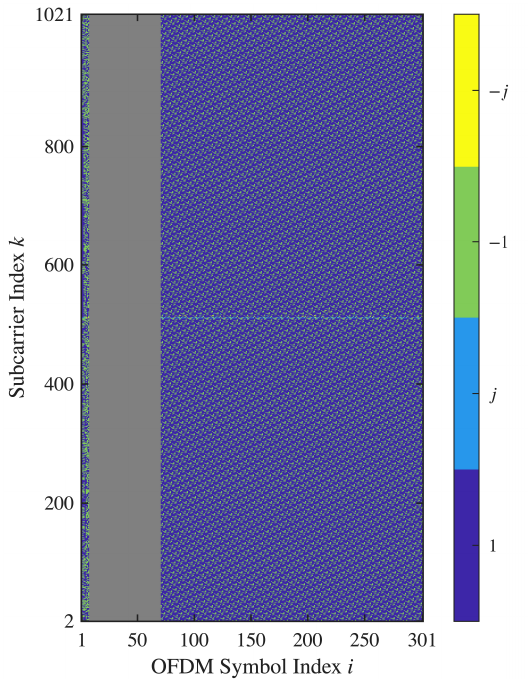}
  \caption{A color-coded plot of $\D_m$ for $m = 814 \in \Mex$.  Readers are
    encouraged to magnify the plot to appreciate its fine structure.  The
    horizontal and vertical axes are ordered respectively by OFDM symbol index
    $i \in \Ido$ and subcarrier index $k \in \Klnp$.  The narrow vertical strip
    to the far left is the deviation within the frame's header,
    $\{\D_{mi} \mid i \in \IHm\}$. The gray region spans the non-QPSK symbols
    for this frame, with indices $i \in \{\Idt \setminus \IQm\}$. The frame's
    T-code tiles the remainder of the frame with a tessellated pattern.  The
    thin horizontal line and speckles near the center ($k\approx511$) are
    decoding artifacts caused by increased phase noise at the far edges of the
    Starlink channel.  Note how the T-code and the header deviation exhibit only
    green and blue colors, consistent with $D_{mik} \in \{1,-1 \}$.}
  \label{fig:frame-d-example}
\end{figure}

Let $\IHm \define \{ i \in \Idt \mid i \le \ihm \}$ denote the header-symbol
indices for the $m$th frame, where $\ihm$ is the index of the final header
symbol.  Fig.~\ref{fig:mhead-hist} shows the distribution of $\ihm$ across the
exemplar frames.  Within the header region, individual OFDM symbols often
manifest long sequences of value $-1$.  These can be seen as vertical blue
strips on close inspection of Fig.~\ref{fig:frame-d-example}. Perhaps certain
header regions are reserved for data only transmitted intermittently and are
otherwise set to $-1$.  In any case, these runs of constant symbols and other
obvious header structure clearly indicate that the header data are not
encrypted.

\paragraph*{T-Codes}
Let $\ITm \define \{ i \in \IQm \mid i > \ihm \}$ denote the indices of
post-header QPSK-modulated OFDM symbols. Inspection of
$\{\D_{mi} \mid i \in \ITm\}$ reveals a tessellated pattern, an example of which
is observable in Fig.~\ref{fig:frame-d-example}. Between successive OFDM
symbols, the deviation's BPSK sequence is circularly shifted by 16 subcarriers
from each OFDM symbol to the next.  Within each OFDM symbol, the deviation's
sequence repeats at 60-subcarrier intervals for $k \in \Klnp$. We refer to the
60-bit BPSK sequence as a tessellation code, or \textit{T-code}. Whereas the
tessellation periodicity is consistent across frames, the T-codes themselves
vary from frame to frame. If non-QPSK OFDM symbols interrupt a continuous
sequence of QPSK-modulated OFDM symbols, the pattern continues as if there were
no interruption.  Continuity is verified by observing that the circular-shift
phase accumulates correctly across such gaps. Let $\NT = 60$ be the T-code
length and $\dT=16$ the circular shift period.  Then for all $i \in \ITm$,
$k \in \Klnp$ we may write
\begin{align}
  \label{eq:tcode60}
  D_{mik} &= D_{mi(k+\NT) \Mod |\Klnp|} \\
  \label{eq:tcode16}
  \D_{m(i+\Delta i)} &= \mathtt{circshift}(\D_{mi},~ \dT \cdot \Delta i)
\end{align}
where $\Delta i$ is the OFDM symbol index difference and
$\mathtt{circshift}(\vb{x},d)$ is a function that circularly shifts elements of
the vector $\vb{x}$ by $d$ positions.  We provide all unique T-codes within the
exemplar frames in the Supplementary Material.

\section{Discussion}
\label{sec:known-predictable}
\subsection{T-Codes:  Significance, Persistence, and Diversity}
\subsubsection{Significance}
The repeating T-code structure implies that the information content of large
blocks within Starlink frames is only $\NT = 60$ bits.  If these bits can be
determined (e.g., by circularly shifting, stacking, and averaging), then the
QPSK-modulated data within each frame can be coherently accumulated. As shown in
Section~\ref{sec:processing-gain}, such extended coherent processing directly
translates into increased processing gain.  Note that T-code symbols appear to
be fairly common: across the exemplar frames, $21.5\%$ of all post-header OFDM
symbols were T-code symbols.

\subsubsection{Persistence}
Whether any given signal feature is useful for opportunistic PNT depends on its
reliability and persistence. We therefore offer several observations and
conjectures regarding the purpose and durability of the low-entropy structures
discovered in Section~\ref{sec:t-codes}.  SpaceX's 2024 patent
\cite{mccormick2024ofdm} indicates that the Starlink SV-side downlink signal
processing chain includes both a single-tap whitening scrambler and a
low-density parity-check (LDPC) encoder.  Whitening scramblers disperse signal
energy to ensure a uniform power spectral density, thereby maintaining DC
balance and reducing the peak-to-average power ratio of the transmitted OFDM
waveform.  The use of a single-tap architecture is significant: by applying a
deterministic complex-domain phase rotation specific to each subcarrier
independently, the transmitter preserves inter-carrier orthogonality and enables
computationally efficient, memoryless descrambling at the receiver.


The presence of a constant reference template is consistent with the output of a
single-tap scrambler, while the observed tessellation pattern is consistent with
the numerological interaction between a 16-wide vectorized hardware processing
architecture and a 60-bit periodicity introduced by structured channel coding
(e.g., an LDPC encoder). The observed frame-to-frame T-code variations suggest
that the scrambler state is re-seeded between frames.

We conjecture that the tesselation pattern shown in
Fig. \ref{fig:frame-d-example}, which is present in some form in the majority of
exemplar frames, is the result of a single-tap whitening scrambler and an LDPC
encoder operating on \emph{unladen} payload allocations---OFDM symbols that are
not bearing any user data. Such empty resource blocks arise due to time-varying
user demand and unavoidable inefficiency in resource scheduling.  Starlink SVs
likely populate empty payload allocations with a deterministic fill pattern
(e.g., an all-ones vector) prior to scrambling and LDPC encoding. The resulting
QPSK symbols thus appear to be user data upon demodulation, yet are in fact
low-entropy (60-bit) transformations of the same reference template.

These observations suggest that T-codes are neither arbitrary artifacts nor
transient byproducts of a particular software release, but are instead tied to
the transmitter's physical-layer architecture. As such, they are likely to
remain a durable feature of the Starlink downlink, even if the fraction of OFDM
symbols exhibiting T-code structure varies with per-beam user demand.

\subsubsection{Diversity}
Although a 60-bit-sequence set could theoretically include $2^{60}$ members,
analysis of T-codes extracted from the exemplar frames suggests a much smaller
active set. Of the 1009 exemplar frames, 530 contained T-codes, among which
there were only 40 unique sequences. Each unique sequence appeared on average
13.25 times, the least frequent 3 times and the most frequent 24 times.  These
results suggest that the active T-code family comprises far fewer than $2^{60}$
unique T-codes.

\subsection{Processing Gain and TOA Bounds}
To understand the implications of these discoveries for PNT, we approximate the
potential processing gain now available via exploitation of all low-entropy
structure in the Starlink frame known to date, relying on empirical data
obtained from the exemplar frames.  For this analysis, we make several
simplifying assumptions.  First, we conservatively ignore the CP on all OFDM
symbols other than the PSS and the SSS.  Second, we treat the header as unknown data.  Third, we assume that
T-code symbols can be distinguished from non-T-code symbols even at low SNR.

To calculate the frame-level processing gain, let $\bar{N}$ denote the average
number of information symbols per frame to which we can assign a non-uniform
distribution.  Since symbols with a uniform prior do not contribute to the
metric $L$, we may substitute $\bar{N}$ for $N$ in
\eqref{eq:processing-gain}. We count in $\bar{N}$ all PSS, SSS, and edge pilot
symbols, as these are fully known.  For frames containing T-codes, we assume
that the corresponding T-code sequence is hard-decoded by stacking, shifting,
and combining the frame's T-code symbols.  Let $|\ITm|$ denote the number of
T-code OFDM symbols in frame $m$, and define the stacking factor for frame $m$
as $M_m = |\ITm||\Klnp|/\NT$.  Define $\bar{M}$ as the average of $M_m$ over the
exemplar frames $m \in \Mex$.  Thus, T-code information symbols are included in
$\bar{N}$ as the hard-decoded bits that permit the assignment of a non-uniform
distribution to each symbol.

Next, we estimate the average (across frames) of the correlation coefficient
$\mu$ from \eqref{eq:processGainCoeffs}.  For each subcarrier $k \in \Kl$, we
construct the local replica element $l_k$ as follows.  For pilot symbols, $l_k$
is set to the known information symbol value, and since the local replica is
exact, $|\mu| = 1$. For T-code symbols, $l_k$ is reconstructed from the
hard-decoded T-code bits and the known tesselation pattern. Assume that $l_k$
agrees with the true symbol with probability $(1 - p_\text{e})$ and disagrees
with probability $p_\text{e}$.  Then $|\mu| = |1 - 2p_\text{e}|$, where
bit-error probability for coherent BPSK demodulation is
\begin{equation*}
  p_\text{e} = \frac{1}{2} \operatorname{erfc}\left(\sqrt{\bar{M}\,\SNRpre}\right)
\end{equation*}
with $\operatorname{erfc}(\cdot)$ being the complementary error function.  The
average correlation coefficient $\bar{\mu}$ is calculated as the weighted
average of $|\mu|$ across all exploitable symbols.

For the exemplar frames, whose mean $\SNRpre$ is $\qty{13.8}{\decibel}$,
$\bar{N} \approx 69{,}500$ and $\bar{\mu} \approx 1$.  Applying
\eqref{eq:processing-gain} yields an average processing gain
$\bar{L} \approx \qty{48.4}{\decibel}$, which is only \qty{6.6}{\decibel} below
the theoretical maximum $L \approx \qty{55}{\decibel}$ calculated in
Section~\ref{sec:processing-gain} for full-frame correlation with completely
known symbols.

At lower SNR, processing gain degrades due to increased T-code bit estimation
errors.  However, for the exemplar frames, the large stacking factor
$\bar{M} \approx 1044$ provides a boost of approximately \qty{30}{\decibel} to
the effective SNR for T-code bit decisions.  Thus, even at
$\SNRpre = \qty{-25}{\decibel}$, T-code recovery remains nearly perfect,
yielding $\bar{\mu} \approx 0.995$ and $\bar{L} \approx \qty{48.4}{\decibel}$.

The gray traces in Fig.~\ref{fig:zz-cr-bounds} show the resulting single-frame
TOA precision bounds when exploiting all known low-entropy elements (LEEs)---the
PSS, SSS, edge pilots, and T-codes---assuming T-code occurrence statistics as
observed in the exemplar frames. The bounds are computed with the receiver's
local oscillator center frequency tuned such that the edge pilots occupy
subcarriers at the outer edges of the captured bandwidth, thereby maximizing the
mean-square bandwidth and tightening both the ZZB and CRB.  Notably, the LEEs
ZZB remains coincident with its corresponding CRB across the entire
pre-correlation SNR range plotted, matching the behavior of the full-frame
bounds. This indicates that the discovered predictable elements are sufficient
to suppress ambiguity sidelobes in the local replica correlation function.

\subsection{TOA Based on Edge Pilots}
\begin{figure}[t]
  \centering
  \includegraphics[width=\columnwidth]{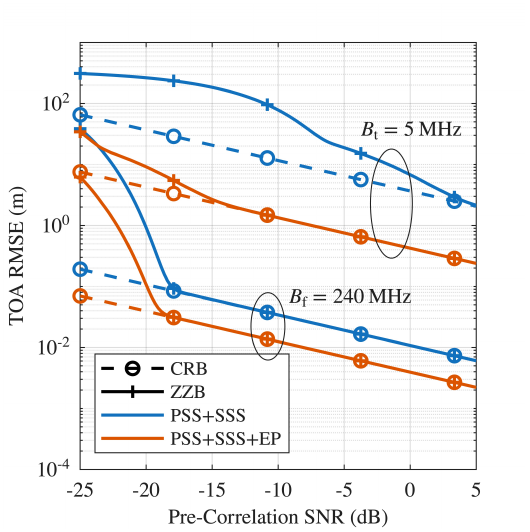}
  \caption{Lower bounds on single-frame TOA RMSE vs. $\SNRpre$ for coherent
    processing with the PSS, SSS, and edge pilots (EP) under two representative
    capture bandwidths. From top to bottom, bounds are shown for (1) narrowband
    correlation with a \qty{5}{\mega\hertz} subband of the PSS+SSS combination;
    (2) narrowband correlation with a coherent combination of the
    \qty{5}{\mega\hertz} PSS+SSS subband concatenated in time with a single edge
    pilot band; (3) full-bandwidth correlation over \qty{240}{\mega\hertz} with
    the PSS+SSS combination, as shown in Fig.~\ref{fig:zz-cr-bounds}; and (4)
    full-bandwidth correlation over \qty{240}{\mega\hertz} with PSS+SSS and both
    edge pilot bands.}
  \label{fig:zz-cr-bounds-edge}
\end{figure}

A lower-performance but especially simple implementation of Starlink-based
opportunistic LEO PNT is possible with only the PSS, SSS, and edge pilots.
Fig.~\ref{fig:zz-cr-bounds-edge} shows TOA RMSE bounds as functions of $\SNRpre$
when the PSS+SSS local replica is augmented with the known edge pilot
symbols. For full-bandwidth processing with
$B_\text{f} = \qty{240}{\mega\hertz}$, adding both edge pilot bands yields only
a modest change in the ZZB-CRB divergence point, but reduces the TOA RMSE bound
by an approximately constant factor of $2.75$ across the plotted SNR range due
to additional processing gain provided by the edge pilots.

For narrowband processing with a tight $B_\text{t} = \qty{5}{\mega\hertz}$,
augmenting the narrowly filtered PSS+SSS with a single edge pilot provides a
larger benefit. Concatenating the edge pilot band with a matched
\qty{5}{\mega\hertz} PSS+SSS subband increases the effective coherent span and
significantly suppresses TOA ambiguity, shifting the ZZB-CRB divergence point
from approximately $6.2$ to \qty{-12.0}{\decibel}.

\subsection{Other Frame-Invariant Data}
Reference \cite{kozhaya2025unveiling} reported additional inter-frame-correlated
data besides the PSS, SSS, header, and edge pilots, modeling these additional
``beacon'' data as appearing deterministically in all frames.
Fig. \ref{fig:qpsk-ratio-corr} does indicate various levels of pairwise
correlation between frames, but as we have revealed, this is due to the
small-family T-code deviations from the reference template $\Tref$.  Since
T-codes vary from frame to frame and are approximately zero-mean, there is no
guarantee of significant inter-frame correlation of QPSK-modulated data unless
one wipes the T-code modulation from the underlying reference template $\Tref$.
We carefully searched multiple Starlink satellite signals for the purported
additional frame-to-frame repeated data using the averaging technique presented
in Section \ref{sec:edge-pilots} and did not find them.  We conjecture that the
additional correlated data found by \cite{kozhaya2025unveiling} are not actually
invariant across Starlink frames but only appeared so in the cases analyzed
therein.

\section{Conclusion}
\label{sec:conclusion}
We developed an acquisition and demodulation framework to hard-decode Starlink
Ku-band downlink frames and applied it to a corpus of frames captured across
multiple satellites and dates.  Our analysis revealed key low-entropy structures
whose existence or details were previously unknown to the public: edge pilots,
the reference template, and T-codes.  Exploiting these discoveries yields
approximately \qty{48}{\decibel} of processing gain, making precise TOA
extraction feasible at minimum SNRs well below those required for
communications. These results support robust opportunistic PNT using compact
feedhorn receivers and suggest that Starlink side-beam signals may be far more
usable for navigation than previously assumed.

\section*{Acknowledgments}
Research was supported by the U.S. Department of Transportation (DOT) under Grant
69A3552348327 for the CARMEN+ University Transportation Center, by
another U.S. DOT Grant for the CARNATIONS University Transportation Center, by the
Department of Defense NDSEG Fellowship program, and by affiliates of the
\verb|6G@UT| center within the Wireless Networking and Communications Group at
The University of Texas at Austin.

\appendices
\section{Edge Pilot Codes}
\label{app:pilot-codes}
The hexadecimal numbers $\qpk$ used to define the edge pilot symbols $\spik$ are
given below for each of the 16 pilot subcarrier indices
$k \in \{488, 489, ..., 495, 528, 529, ..., 535\}$.  The edge pilots are also
given directly in the Supplementary Material as a $|\Kp|$-by-$|\Idt|$
complex-valued matrix whose $(k,i)$th element is $X_{mik}$ from
\eqref{eq:pilotCoeff}.

\begin{equation}
  \notag
  \begin{array}{l}
    q_{p488} = \\
    \quad
    \begin{array}{rrrrrrrr}
      & &  \text{\texttt{76}} & \text{\texttt{3404}} & \text{\texttt{6DA4}} & \text{\texttt{5F89}} & \text{\texttt{042D}} & \text{\texttt{0117}}\\
      \text{\texttt{E163}} & \text{\texttt{167D}} & \text{\texttt{4AE8}} & \text{\texttt{32D8}} & \text{\texttt{5751}} & \text{\texttt{5F3C}} & \text{\texttt{AD90}} & \text{\texttt{3376}}\\
      \text{\texttt{97FB}} & \text{\texttt{8F1C}} & \text{\texttt{D048}} & \text{\texttt{EFEF}} & \text{\texttt{559E}} & \text{\texttt{CD79}} & \text{\texttt{688B}} & \text{\texttt{CBBF}}\\
      \text{\texttt{44D2}} & \text{\texttt{FA9B}} & \text{\texttt{DFAE}} & \text{\texttt{639D}} & \text{\texttt{B5D7}} & \text{\texttt{B1DD}} & \text{\texttt{2DDC}} & \text{\texttt{E4EC}}\\
      \text{\texttt{9733}} & \text{\texttt{C0D4}} & \text{\texttt{DCCF}} & \text{\texttt{3172}} & \text{\texttt{A0EC}} & \text{\texttt{34CC}} & \text{\texttt{226C}} & \text{\texttt{530E}}
    \end{array}
  \end{array}	
\end{equation}

\begin{equation}
  \notag
  \begin{array}{l}
    q_{p489} = \\
    \quad
    \begin{array}{rrrrrrrr}
      & &  \text{\texttt{CD}} & \text{\texttt{9AFA}} & \text{\texttt{A654}} & \text{\texttt{147A}} & \text{\texttt{5FE2}} & \text{\texttt{B407}}\\
      \text{\texttt{B51F}} & \text{\texttt{FE15}} & \text{\texttt{215B}} & \text{\texttt{3A71}} & \text{\texttt{6241}} & \text{\texttt{3961}} & \text{\texttt{9628}} & \text{\texttt{A9C3}}\\
      \text{\texttt{3E8E}} & \text{\texttt{3A32}} & \text{\texttt{E514}} & \text{\texttt{6C09}} & \text{\texttt{BD3E}} & \text{\texttt{E902}} & \text{\texttt{6CA5}} & \text{\texttt{2032}}\\
      \text{\texttt{D7FD}} & \text{\texttt{3896}} & \text{\texttt{0FFC}} & \text{\texttt{5259}} & \text{\texttt{9E9B}} & \text{\texttt{8A7F}} & \text{\texttt{6942}} & \text{\texttt{334B}}\\
      \text{\texttt{D4C6}} & \text{\texttt{D99D}} & \text{\texttt{4331}} & \text{\texttt{DEF5}} & \text{\texttt{6745}} & \text{\texttt{70B2}} & \text{\texttt{45FB}} & \text{\texttt{B25F}}
    \end{array}
  \end{array}	
\end{equation}

\begin{equation}
  \notag
  \begin{array}{l}
    q_{p490} = \\
    \quad
    \begin{array}{rrrrrrrr}
      & &  \text{\texttt{02}} & \text{\texttt{481A}} & \text{\texttt{2B27}} & \text{\texttt{8B88}} & \text{\texttt{F096}} & \text{\texttt{C8D1}}\\
      \text{\texttt{74D3}} & \text{\texttt{69D0}} & \text{\texttt{CF67}} & \text{\texttt{81B7}} & \text{\texttt{0EBD}} & \text{\texttt{402D}} & \text{\texttt{6A6F}} & \text{\texttt{4C98}}\\
      \text{\texttt{5DA6}} & \text{\texttt{2658}} & \text{\texttt{66A8}} & \text{\texttt{374D}} & \text{\texttt{C0B3}} & \text{\texttt{E491}} & \text{\texttt{7146}} & \text{\texttt{FE32}}\\
      \text{\texttt{74CA}} & \text{\texttt{5D61}} & \text{\texttt{C3F9}} & \text{\texttt{A31C}} & \text{\texttt{B812}} & \text{\texttt{5F29}} & \text{\texttt{1155}} & \text{\texttt{CBD4}}\\
      \text{\texttt{F4F8}} & \text{\texttt{4E93}} & \text{\texttt{C0D8}} & \text{\texttt{54BB}} & \text{\texttt{C54E}} & \text{\texttt{E144}} & \text{\texttt{43EC}} & \text{\texttt{2DF8}}
    \end{array}
  \end{array}	
\end{equation}

\begin{equation}
  \notag
  \begin{array}{l}
    q_{p491} = \\
    \quad
    \begin{array}{rrrrrrrr}
      & &  \text{\texttt{D8}} & \text{\texttt{DC99}} & \text{\texttt{C265}} & \text{\texttt{4265}} & \text{\texttt{B8C3}} & \text{\texttt{2450}}\\
      \text{\texttt{114C}} & \text{\texttt{37E2}} & \text{\texttt{B725}} & \text{\texttt{A822}} & \text{\texttt{F105}} & \text{\texttt{4B46}} & \text{\texttt{F272}} & \text{\texttt{8771}}\\
      \text{\texttt{22E4}} & \text{\texttt{7109}} & \text{\texttt{F113}} & \text{\texttt{D59E}} & \text{\texttt{37DF}} & \text{\texttt{F418}} & \text{\texttt{FEA3}} & \text{\texttt{627C}}\\
      \text{\texttt{7A5C}} & \text{\texttt{C0A9}} & \text{\texttt{3ABA}} & \text{\texttt{0F94}} & \text{\texttt{08E9}} & \text{\texttt{58DF}} & \text{\texttt{4179}} & \text{\texttt{C4DE}}\\
      \text{\texttt{40CE}} & \text{\texttt{F842}} & \text{\texttt{D333}} & \text{\texttt{632B}} & \text{\texttt{3E77}} & \text{\texttt{BEB3}} & \text{\texttt{4B2E}} & \text{\texttt{6045}}
    \end{array}
  \end{array}	
\end{equation}

\begin{equation}
  \notag
  \begin{array}{l}
    q_{p492} = \\
    \quad
    \begin{array}{rrrrrrrr}
      & &  \text{\texttt{3C}} & \text{\texttt{C5CA}} & \text{\texttt{83B0}} & \text{\texttt{D330}} & \text{\texttt{89B1}} & \text{\texttt{4C3B}}\\
      \text{\texttt{6AC3}} & \text{\texttt{D194}} & \text{\texttt{6359}} & \text{\texttt{726B}} & \text{\texttt{4966}} & \text{\texttt{B2E9}} & \text{\texttt{66BE}} & \text{\texttt{6112}}\\
      \text{\texttt{4A5D}} & \text{\texttt{53E2}} & \text{\texttt{2A73}} & \text{\texttt{EDBE}} & \text{\texttt{B383}} & \text{\texttt{A92F}} & \text{\texttt{06CA}} & \text{\texttt{6CAA}}\\
      \text{\texttt{8A5B}} & \text{\texttt{1ECE}} & \text{\texttt{6954}} & \text{\texttt{6514}} & \text{\texttt{5E28}} & \text{\texttt{6EEE}} & \text{\texttt{1804}} & \text{\texttt{CD79}}\\
      \text{\texttt{A00C}} & \text{\texttt{84FC}} & \text{\texttt{80C8}} & \text{\texttt{7DE9}} & \text{\texttt{DF57}} & \text{\texttt{2F9B}} & \text{\texttt{54AE}} & \text{\texttt{798B}}
    \end{array}
  \end{array}	
\end{equation}

\begin{equation}
  \notag
  \begin{array}{l}
    q_{p493} = \\
    \quad
    \begin{array}{rrrrrrrr}
      & &  \text{\texttt{C7}} & \text{\texttt{7BD5}} & \text{\texttt{9D15}} & \text{\texttt{C2C9}} & \text{\texttt{17EE}} & \text{\texttt{C97F}}\\
      \text{\texttt{B479}} & \text{\texttt{B9F0}} & \text{\texttt{B2BF}} & \text{\texttt{5D2E}} & \text{\texttt{CCD8}} & \text{\texttt{0248}} & \text{\texttt{D2AC}} & \text{\texttt{68C8}}\\
      \text{\texttt{4CEA}} & \text{\texttt{11BA}} & \text{\texttt{D18D}} & \text{\texttt{9F6F}} & \text{\texttt{31B6}} & \text{\texttt{AFD7}} & \text{\texttt{8334}} & \text{\texttt{7943}}\\
      \text{\texttt{562E}} & \text{\texttt{2C68}} & \text{\texttt{32EA}} & \text{\texttt{7682}} & \text{\texttt{8FCD}} & \text{\texttt{AB31}} & \text{\texttt{EFF6}} & \text{\texttt{A9A8}}\\
      \text{\texttt{8EA4}} & \text{\texttt{8E3A}} & \text{\texttt{FA62}} & \text{\texttt{5B2F}} & \text{\texttt{CDA7}} & \text{\texttt{B99B}} & \text{\texttt{0295}} & \text{\texttt{E926}}
    \end{array}
  \end{array}	
\end{equation}

\begin{equation}
  \notag
  \begin{array}{l}
    q_{p494} = \\
    \quad
    \begin{array}{rrrrrrrr}
      & &  \text{\texttt{61}} & \text{\texttt{52EF}} & \text{\texttt{153B}} & \text{\texttt{8511}} & \text{\texttt{0FB0}} & \text{\texttt{B7E2}}\\
      \text{\texttt{4D83}} & \text{\texttt{34B1}} & \text{\texttt{C419}} & \text{\texttt{6DE8}} & \text{\texttt{72B5}} & \text{\texttt{9876}} & \text{\texttt{7BC3}} & \text{\texttt{CB4A}}\\
      \text{\texttt{4827}} & \text{\texttt{A09D}} & \text{\texttt{924A}} & \text{\texttt{A7F5}} & \text{\texttt{7EB9}} & \text{\texttt{46F1}} & \text{\texttt{981D}} & \text{\texttt{036E}}\\
      \text{\texttt{3001}} & \text{\texttt{934B}} & \text{\texttt{10C9}} & \text{\texttt{E22A}} & \text{\texttt{BB6A}} & \text{\texttt{F1F0}} & \text{\texttt{47B3}} & \text{\texttt{A874}}\\
      \text{\texttt{CA95}} & \text{\texttt{E68C}} & \text{\texttt{BA67}} & \text{\texttt{063F}} & \text{\texttt{605F}} & \text{\texttt{D05D}} & \text{\texttt{532A}} & \text{\texttt{AD3C}}
    \end{array}
  \end{array}	
\end{equation}

\begin{equation}
  \notag
  \begin{array}{l}
    q_{p495} = \\
    \quad
    \begin{array}{rrrrrrrr}
      & &  \text{\texttt{CD}} & \text{\texttt{8CAC}} & \text{\texttt{F9DE}} & \text{\texttt{FACD}} & \text{\texttt{2CB9}} & \text{\texttt{8114}}\\
      \text{\texttt{39D8}} & \text{\texttt{B7E1}} & \text{\texttt{6F9E}} & \text{\texttt{09BE}} & \text{\texttt{D473}} & \text{\texttt{7020}} & \text{\texttt{7150}} & \text{\texttt{A86D}}\\
      \text{\texttt{FE24}} & \text{\texttt{EA12}} & \text{\texttt{98CC}} & \text{\texttt{B090}} & \text{\texttt{7F5B}} & \text{\texttt{AB67}} & \text{\texttt{D466}} & \text{\texttt{0462}}\\
      \text{\texttt{C6B1}} & \text{\texttt{0F74}} & \text{\texttt{B8D9}} & \text{\texttt{FA7B}} & \text{\texttt{6F9E}} & \text{\texttt{C139}} & \text{\texttt{9B30}} & \text{\texttt{B43A}}\\
      \text{\texttt{F622}} & \text{\texttt{A894}} & \text{\texttt{B222}} & \text{\texttt{0B6B}} & \text{\texttt{509A}} & \text{\texttt{84AA}} & \text{\texttt{BB58}} & \text{\texttt{D023}}
    \end{array}
  \end{array}	
\end{equation}

\begin{equation}
  \notag
  \begin{array}{l}
    q_{p528} = \\
    \quad
    \begin{array}{rrrrrrrr}
      & &  \text{\texttt{CC}} & \text{\texttt{BF3A}} & \text{\texttt{1692}} & \text{\texttt{9836}} & \text{\texttt{160C}} & \text{\texttt{EC6E}}\\
      \text{\texttt{B741}} & \text{\texttt{7AE6}} & \text{\texttt{C37D}} & \text{\texttt{C1E8}} & \text{\texttt{28CE}} & \text{\texttt{FB60}} & \text{\texttt{CE0E}} & \text{\texttt{6C3B}}\\
      \text{\texttt{546A}} & \text{\texttt{76B0}} & \text{\texttt{AE1E}} & \text{\texttt{7BC0}} & \text{\texttt{E957}} & \text{\texttt{7528}} & \text{\texttt{B0F7}} & \text{\texttt{8F82}}\\
      \text{\texttt{A410}} & \text{\texttt{4EA2}} & \text{\texttt{C316}} & \text{\texttt{B945}} & \text{\texttt{D385}} & \text{\texttt{200C}} & \text{\texttt{7E5A}} & \text{\texttt{1C5B}}\\
      \text{\texttt{48F5}} & \text{\texttt{F9F9}} & \text{\texttt{AF5C}} & \text{\texttt{4BA9}} & \text{\texttt{20AC}} & \text{\texttt{A3A5}} & \text{\texttt{99DB}} & \text{\texttt{9974}}
    \end{array}
  \end{array}	
\end{equation}

\begin{equation}
  \notag
  \begin{array}{l}
    q_{p529} = \\
    \quad
    \begin{array}{rrrrrrrr}
      & &  \text{\texttt{9C}} & \text{\texttt{F72F}} & \text{\texttt{5F5B}} & \text{\texttt{95CE}} & \text{\texttt{7342}} & \text{\texttt{C925}}\\
      \text{\texttt{CF1A}} & \text{\texttt{AF45}} & \text{\texttt{7F18}} & \text{\texttt{2C32}} & \text{\texttt{810E}} & \text{\texttt{2F74}} & \text{\texttt{8670}} & \text{\texttt{5D5F}}\\
      \text{\texttt{A2D9}} & \text{\texttt{C892}} & \text{\texttt{3B01}} & \text{\texttt{73FB}} & \text{\texttt{206B}} & \text{\texttt{4604}} & \text{\texttt{5C6F}} & \text{\texttt{162B}}\\
      \text{\texttt{B9FF}} & \text{\texttt{D051}} & \text{\texttt{DB5E}} & \text{\texttt{5900}} & \text{\texttt{EFD2}} & \text{\texttt{DE24}} & \text{\texttt{D4BB}} & \text{\texttt{3FE8}}\\
      \text{\texttt{7DD7}} & \text{\texttt{76F0}} & \text{\texttt{0B56}} & \text{\texttt{13A7}} & \text{\texttt{D22B}} & \text{\texttt{2821}} & \text{\texttt{E139}} & \text{\texttt{A599}}
    \end{array}
  \end{array}	
\end{equation}

\begin{equation}
  \notag
  \begin{array}{l}
    q_{p530} = \\
    \quad
    \begin{array}{rrrrrrrr}
      & &  \text{\texttt{29}} & \text{\texttt{6319}} & \text{\texttt{D723}} & \text{\texttt{2101}} & \text{\texttt{8995}} & \text{\texttt{3BB7}}\\
      \text{\texttt{30DC}} & \text{\texttt{6046}} & \text{\texttt{E4EC}} & \text{\texttt{5FB4}} & \text{\texttt{8F97}} & \text{\texttt{18D5}} & \text{\texttt{B600}} & \text{\texttt{A015}}\\
      \text{\texttt{78CA}} & \text{\texttt{C315}} & \text{\texttt{9B58}} & \text{\texttt{EE8A}} & \text{\texttt{3066}} & \text{\texttt{6392}} & \text{\texttt{1FBE}} & \text{\texttt{78EE}}\\
      \text{\texttt{7C1E}} & \text{\texttt{8E04}} & \text{\texttt{9B42}} & \text{\texttt{30A1}} & \text{\texttt{4EB4}} & \text{\texttt{9549}} & \text{\texttt{33AB}} & \text{\texttt{64F6}}\\
      \text{\texttt{7B39}} & \text{\texttt{6DD6}} & \text{\texttt{DB12}} & \text{\texttt{BCBB}} & \text{\texttt{3CCA}} & \text{\texttt{60EA}} & \text{\texttt{79E0}} & \text{\texttt{614B}}
    \end{array}
  \end{array}	
\end{equation}

\begin{equation}
  \notag
  \begin{array}{l}
    q_{p531} = \\
    \quad
    \begin{array}{rrrrrrrr}
      & &  \text{\texttt{10}} & \text{\texttt{17FB}} & \text{\texttt{BD3D}} & \text{\texttt{0398}} & \text{\texttt{1EE9}} & \text{\texttt{F442}}\\
      \text{\texttt{4D47}} & \text{\texttt{3B8A}} & \text{\texttt{73E1}} & \text{\texttt{36C7}} & \text{\texttt{7795}} & \text{\texttt{6EAE}} & \text{\texttt{BD4C}} & \text{\texttt{A51E}}\\
      \text{\texttt{9B70}} & \text{\texttt{D9F5}} & \text{\texttt{D106}} & \text{\texttt{57F2}} & \text{\texttt{6859}} & \text{\texttt{5A5C}} & \text{\texttt{3687}} & \text{\texttt{D2DD}}\\
      \text{\texttt{06C9}} & \text{\texttt{8630}} & \text{\texttt{F817}} & \text{\texttt{CABE}} & \text{\texttt{F3EE}} & \text{\texttt{6608}} & \text{\texttt{2235}} & \text{\texttt{0A70}}\\
      \text{\texttt{F10A}} & \text{\texttt{29A8}} & \text{\texttt{7402}} & \text{\texttt{12A9}} & \text{\texttt{CF7E}} & \text{\texttt{7D81}} & \text{\texttt{4D60}} & \text{\texttt{A69C}}
    \end{array}
  \end{array}	
\end{equation}

\begin{equation}
  \notag
  \begin{array}{l}
    q_{p532} = \\
    \quad
    \begin{array}{rrrrrrrr}
      & &  \text{\texttt{71}} & \text{\texttt{2EA4}} & \text{\texttt{82B2}} & \text{\texttt{8E96}} & \text{\texttt{676E}} & \text{\texttt{65D0}}\\
      \text{\texttt{9994}} & \text{\texttt{9655}} & \text{\texttt{8731}} & \text{\texttt{4F2B}} & \text{\texttt{562D}} & \text{\texttt{0E75}} & \text{\texttt{0FE5}} & \text{\texttt{66E8}}\\
      \text{\texttt{9205}} & \text{\texttt{A8D4}} & \text{\texttt{DFED}} & \text{\texttt{2C4F}} & \text{\texttt{AFFC}} & \text{\texttt{5ED1}} & \text{\texttt{EA6F}} & \text{\texttt{B63E}}\\
      \text{\texttt{C135}} & \text{\texttt{1344}} & \text{\texttt{4006}} & \text{\texttt{B78A}} & \text{\texttt{DFB4}} & \text{\texttt{BDB6}} & \text{\texttt{CB05}} & \text{\texttt{4706}}\\
      \text{\texttt{01C9}} & \text{\texttt{F8F4}} & \text{\texttt{9014}} & \text{\texttt{2306}} & \text{\texttt{9C9F}} & \text{\texttt{BD68}} & \text{\texttt{D292}} & \text{\texttt{C16F}}
    \end{array}
  \end{array}	
\end{equation}

\begin{equation}
  \notag
  \begin{array}{l}
    q_{p533} = \\
    \quad
    \begin{array}{rrrrrrrr}
      & &  \text{\texttt{58}} & \text{\texttt{4E9F}} & \text{\texttt{48AC}} & \text{\texttt{A087}} & \text{\texttt{84E6}} & \text{\texttt{9664}}\\
      \text{\texttt{4C78}} & \text{\texttt{ED96}} & \text{\texttt{84FC}} & \text{\texttt{484F}} & \text{\texttt{32AA}} & \text{\texttt{1B4D}} & \text{\texttt{A8E9}} & \text{\texttt{5457}}\\
      \text{\texttt{358D}} & \text{\texttt{F89F}} & \text{\texttt{E8B9}} & \text{\texttt{D84D}} & \text{\texttt{47F3}} & \text{\texttt{0D3C}} & \text{\texttt{A2F2}} & \text{\texttt{DDF0}}\\
      \text{\texttt{E76E}} & \text{\texttt{57F1}} & \text{\texttt{4A44}} & \text{\texttt{6753}} & \text{\texttt{26ED}} & \text{\texttt{CF15}} & \text{\texttt{052C}} & \text{\texttt{B62B}}\\
      \text{\texttt{7DF0}} & \text{\texttt{EBE6}} & \text{\texttt{2305}} & \text{\texttt{7605}} & \text{\texttt{CF24}} & \text{\texttt{06E2}} & \text{\texttt{5BD5}} & \text{\texttt{6B3B}}
    \end{array}
  \end{array}	
\end{equation}

\begin{equation}
  \notag
  \begin{array}{l}
    q_{p534} = \\
    \quad
    \begin{array}{rrrrrrrr}
      & &  \text{\texttt{4A}} & \text{\texttt{F2EC}} & \text{\texttt{F329}} & \text{\texttt{83A9}} & \text{\texttt{E781}} & \text{\texttt{852F}}\\
      \text{\texttt{6E90}} & \text{\texttt{DC6C}} & \text{\texttt{CE90}} & \text{\texttt{1863}} & \text{\texttt{F527}} & \text{\texttt{E038}} & \text{\texttt{DA22}} & \text{\texttt{C0CE}}\\
      \text{\texttt{02E4}} & \text{\texttt{4FA0}} & \text{\texttt{5637}} & \text{\texttt{18D9}} & \text{\texttt{3E74}} & \text{\texttt{5429}} & \text{\texttt{3962}} & \text{\texttt{B435}}\\
      \text{\texttt{94CC}} & \text{\texttt{2EE4}} & \text{\texttt{27FA}} & \text{\texttt{E6F1}} & \text{\texttt{5C12}} & \text{\texttt{38D9}} & \text{\texttt{C85A}} & \text{\texttt{BC4E}}\\
      \text{\texttt{303F}} & \text{\texttt{3AEC}} & \text{\texttt{3404}} & \text{\texttt{A523}} & \text{\texttt{10CA}} & \text{\texttt{C037}} & \text{\texttt{8665}} & \text{\texttt{E19A}}
    \end{array}
  \end{array}	
\end{equation}

\begin{equation}
  \notag
  \begin{array}{l}
    q_{p535} = \\
    \quad
    \begin{array}{rrrrrrrr}
      & &  \text{\texttt{08}} & \text{\texttt{4AA7}} & \text{\texttt{3DF9}} & \text{\texttt{F605}} & \text{\texttt{3582}} & \text{\texttt{9A71}}\\
      \text{\texttt{6EC9}} & \text{\texttt{4D95}} & \text{\texttt{AA69}} & \text{\texttt{01B4}} & \text{\texttt{1E81}} & \text{\texttt{AEF2}} & \text{\texttt{8B03}} & \text{\texttt{F08C}}\\
      \text{\texttt{DE7D}} & \text{\texttt{4542}} & \text{\texttt{5B11}} & \text{\texttt{6400}} & \text{\texttt{9D56}} & \text{\texttt{459C}} & \text{\texttt{4286}} & \text{\texttt{E269}}\\
      \text{\texttt{F4B8}} & \text{\texttt{EBDB}} & \text{\texttt{A8BF}} & \text{\texttt{6FC7}} & \text{\texttt{9847}} & \text{\texttt{B08A}} & \text{\texttt{69F7}} & \text{\texttt{9AF6}}\\
      \text{\texttt{E6A7}} & \text{\texttt{AF05}} & \text{\texttt{DA50}} & \text{\texttt{4455}} & \text{\texttt{BA72}} & \text{\texttt{727D}} & \text{\texttt{D7BE}} & \text{\texttt{7744}}
    \end{array}
  \end{array}	
\end{equation}

\bibliographystyle{IEEEtran} 
\bibliography{pangea}

\end{document}